\documentclass[final,authoryear,5p,times,twocolumn]{elsarticle}
\usepackage{graphicx}
\usepackage{amssymb}
\usepackage{natbib}

\journal{New Astronomy}

\newcommand{\mnras}{MNRAS}
\newcommand{\apj}{ApJ}
\newcommand{\apjl}{ApJ}
\newcommand{\aj}{AJ}
\newcommand{\apjs}{ApJS}
\newcommand{\aap}{A\&A}
\newcommand{\araa}{ARA\&A}
\newcommand{\nat}{Nature}

\newcommand{\physrep}{Phys.~Rep.}
\newcommand{\arcsec}{''}

\begin{document}

\begin{frontmatter}

\title{The Environments of Short-Duration Gamma-Ray Bursts and
Implications for their Progenitors}

\author[har]{Edo Berger\corref{cor1}}

\cortext[cor1]{Corresponding author}

\address[har]{Harvard-Smithsonian Center for Astrophysics, 60 Garden
Street, Cambridge, MA 02139, USA}

\begin{abstract} The study of short-duration gamma-ray bursts (GRBs)
experienced a complete revolution in recent years thanks to the
discovery of the first afterglows and host galaxies starting in May
2005.  These observations demonstrated that short GRBs are
cosmological in origin, reside in both star forming and elliptical
galaxies, are not associated with supernovae, and span a wide
isotropic-equivalent energy range of $\sim 10^{48}-10^{52}$ erg.
However, a fundamental question remains unanswered: What are the
progenitors of short GRBs?  The most popular theoretical model invokes
the coalescence of compact object binaries with neutron star and/or
black hole constituents.  However, additional possibilities exist,
including magnetars formed through prompt channels (massive star
core-collapse) and delayed channels (binary white dwarf mergers, white
dwarf accretion-induced collapse), or accretion-induced collapse of
neutron stars.  In this review I summarize our current knowledge of
the galactic and sub-galactic environments of short GRBs, and use
these observations to draw inferences about the progenitor population.
The most crucial results are: (i) some short GRBs explode in dead
elliptical galaxies; (ii) the majority of short GRBs occur in star
forming galaxies; (iii) the star forming hosts of short GRBs are
distinct from those of long GRBs, and instead appear to be drawn from
the general field galaxy population; (iv) the physical offsets of
short GRBs relative to their host galaxy centers are significantly
larger than for long GRBs; (v) there is tentative evidence for large
offsets from short GRBs with optical afterglows and no coincident
hosts; (vi) the observed offset distribution is in good agreement with
predictions for NS-NS binary mergers; and (vii) short GRBs trace
under-luminous locations within their hosts, but appear to be more
closely correlated with the rest-frame optical light (old stars) than
the UV light (young massive stars).  Taken together, these
observations suggest that short GRB progenitors belong to an old
stellar population with a wide age distribution, and generally track
stellar mass.  These results are fully consistent with NS-NS binary
mergers and rule out a dominant population of prompt magnetars.
However, a partial contribution from delayed magnetar formation or
accretion-induced collapse is also consistent with the data.
\end{abstract}


\end{frontmatter}

\section{Introduction}
\label{sec:intro}

Gamma-ray bursts (GRBs) are short, intense and non-repeating flashes
of $\gamma$-ray radiation originating at cosmological distances.
While GRBs exhibit a broad diversity in their prompt $\gamma$-ray
emission (e.g., duration, spectral shape, peak energy, brightness),
they can still be divided into two basic categories: short-duration
and long-duration with a separation at about 2 sec \citep{kmf+93}.
The short GRBs have durations as short as $\sim 10$ msec, while the
long events extend to hundreds of seconds.  In addition, short GRBs
tend to exhibit harder $\gamma$-ray spectra than the long-duration
events, and generally have a lower fluences.

The basic bimodality of GRB durations provided an early clue that the
progenitors of the two classes are likely to be distinct.  Within the
broad range of possible scenarios, two popular models have emerged:
The collapse of rapidly-rotating massive stars (``collapsars'';
\citep{mw99}) in the case of long GRBs, and the coalescence of compact
object binaries (with neutron star and/or black hole constituents --
NS-NS/NS-BH; \citep{elp+89,pac91,npp92}) in the case of short GRBs.
The key attractions of this mapping are the potential for a large
energy release from both progenitor classes, and the expected typical
timescale for each progenitor: A free-fall timescale of $t_{ff}\approx
30\,{\rm s}\,(M/10\,M_\odot)^{-1/2}(R/10^{10}\,{\rm cm})^{3/2}$ for
collapsars, and a dynamical timescale of milliseconds for the compact
merger remnants of neutron stars and black holes.  However, other
progenitor systems have also been proposed for short GRBs, for example
magnetars, thought to be the power source behind soft $\gamma$-ray
repeaters \citep{td95}, accretion-induced collapse (AIC) of neutron
stars \citep{qwc+98}, and delayed magnetar formation through binary
white dwarf mergers or white dwarf AIC \citep{lwc+06,mqt08}.

Until a decade ago, the distances, energy scale, geometry,
environments, and progenitors of GRBs, as well as the relation between
the two burst classes, remained uncertain due to the lack of precise
positions.  The discovery of long-wavelength, long-lived
``afterglows'' from long GRBs in 1997 provided the first glimpse at
these properties \citep{cfh+97,fkn+97,vgg+97}.  Indeed, the
sub-arcsecond positions enabled by long GRB afterglow detections
demonstrated a cosmological origin \citep{mdk+97}, an energy scale of
$\sim 10^{51}$ erg \citep{fks+01,bkf03,bkp+03,bfk03}, significant
collimation ($\sim 10^\circ$ jets; \citep{hbf+99,sgk+99}), and direct
evidence for relativistic expansion \citep{tfb+04}.  Intense afterglow
and host galaxy observations also linked long GRBs with the deaths of
massive stars, mainly through their exclusive location in star forming
galaxies (e.g., \citep{bdk+98,dkb+98,ftm+99}), their strong
correlation with the rest-frame ultraviolet (UV) light of their hosts
\citep{bkd02,fls+06}, and their association with Type Ic core-collapse
supernovae \citep{hsm+03,smg+03}.

The afterglows of short GRBs were discovered only in 2005.  Prior to
that point only a few afterglow searches were possible due to the
relative faintness of the $\gamma$-ray emission, and hence a low event
rate and large and delayed error circles \citep{hbc+02}.  In
retrospect, these searches were woefully inadequate, reaching only
about 21 mag at $\delta t\approx 1$ day in the optical and $\sim 0.5$
mJy at $\delta t\approx {\rm few}$ days in the radio.  The launch of
NASA's {\it Swift} satellite in late 2004, provided the first chance
for rapid and accurate positions for short GRBs, and indeed led to the
detection of the first X-ray \citep{gso+05,bpp+06}, optical
\citep{ffp+05,hwf+05}, and radio \citep{bpc+05} afterglows in 2005.

As in the case of long GRBs, the determination of accurate positions
revolutionized the study of short GRBs.  First, it led to an
association with galaxies at cosmological distances (e.g.,
\citep{bpc+05,ffp+05,hwf+05,bpp+06,pbc+06,bfp+07}) and hence an energy
scale of $\sim 10^{49}-10^{52}$ erg (assuming isotropy)
\citep{ber07,nak07}.  Second, it led to the association of some events
with elliptical galaxies pointing to an old progenitor population
\citep{bpc+05,gso+05,bpp+06}.  Third, it demonstrated that the
afterglow emission is similar to that of long GRBs, albeit with a
generally lower luminosity \citep{bpc+05}.  Finally, it provided a
rough estimate of the short GRB event rate \citep{ngf06}.

{\it Despite these fundamental results the progenitors of short GRBs
remain unidentified at the present.}  The key observational test of
the NS-NS/NS-BH merger model, the detection of coincident
gravitational waves, is at least several years away.  Similarly,
theoretical predictions of early optical/UV emission from an
accompanying ``mini-supernova'' \citep{lp98}, caused by the ejection
of radioactive material from the merging system, are highly uncertain,
and even the most optimistic predictions lead to a faint and
rapidly-fading signal that may be challenging to detect.  Thus, the
most promising avenue for progress at the present comes from
statistical studies of the environments of short GRBs, both on
galactic and sub-galactic scales (e.g., \citep{pbc+06,ber09,fbf10}).
These studies benefit from many of the same techniques that linked
long GRBs with the death of massive stars, and allow for comparison
with theoretical predictions.

In this review, I present the current state of our knowledge about the
redshift distribution of short GRBs, the demographics and detailed
properties of their host galaxies, and their locations within their
hosts.  The structure of this review is as follows.  In
\S\ref{sec:disc} I summarize the discovery of short GRB afterglows,
and the subsequent identifications of their host galaxies and
redshifts; the detailed properties of the hosts (luminosities,
metallicities, star formation rates, masses, stellar population ages)
are discussed in \S\ref{sec:hosts} and \S\ref{sec:hosts2}; in
\S\ref{sec:hst} I discuss the sub-galactic environments of short GRBs,
utilizing mainly high-resolution {\it Hubble Space Telescope}
observations; I discuss the possibility of large progenitor offsets
(due to kicks or globular cluster origin) in \S\ref{sec:kicks}; and
finally, in \S\ref{sec:conc} I use these results to place constraints
on the progenitor population.

\section{The Discovery of Short GRB Afterglows, Host Galaxies, and 
Redshifts}
\label{sec:disc}

The discovery of short GRB afterglows starting in May 2005 led to the
first identifications of their host galaxies and hence to distance
measurements.  The first short GRB with an afterglow detection,
GRB\,050509B, was localized to a positional accuracy of about $5''$
with the {\it Swift} X-ray Telescope (XRT) \citep{gso+05,bpp+06}.  No
optical or radio emission was detected.  The X-ray error circle
appeared to coincide with the outskirts of an elliptical galaxy at
$z=0.226$ \citep{gso+05,hsg+05,bpp+06}, with a probability of chance
coincidence of $\sim 10^{-3}$.  However, the error circle contained
additional fainter galaxies possibly at higher redshift.

Only two months later, GRB\,050709 was the first short burst localized
to sub-arcsecond precision through the detection of X-ray (with {\it
Chandra}) and optical emission \citep{ffp+05,hwf+05}.  The resulting
afterglow position coincided with the outskirts of an irregular star
forming galaxy at $z=0.161$ \citep{ffp+05,cmi+06}.  Despite the
on-going star formation activity within the host galaxy, the burst was
not accompanied by a supernova explosion, indicating that the
progenitor was not likely to be a massive star \citep{ffp+05,hwf+05}.
However, due to the presence of active star formation, an association
with a young progenitor system such as a magnetar could not be
excluded.

\begin{figure}[ht!]
\centerline{\includegraphics[width=3.5in]{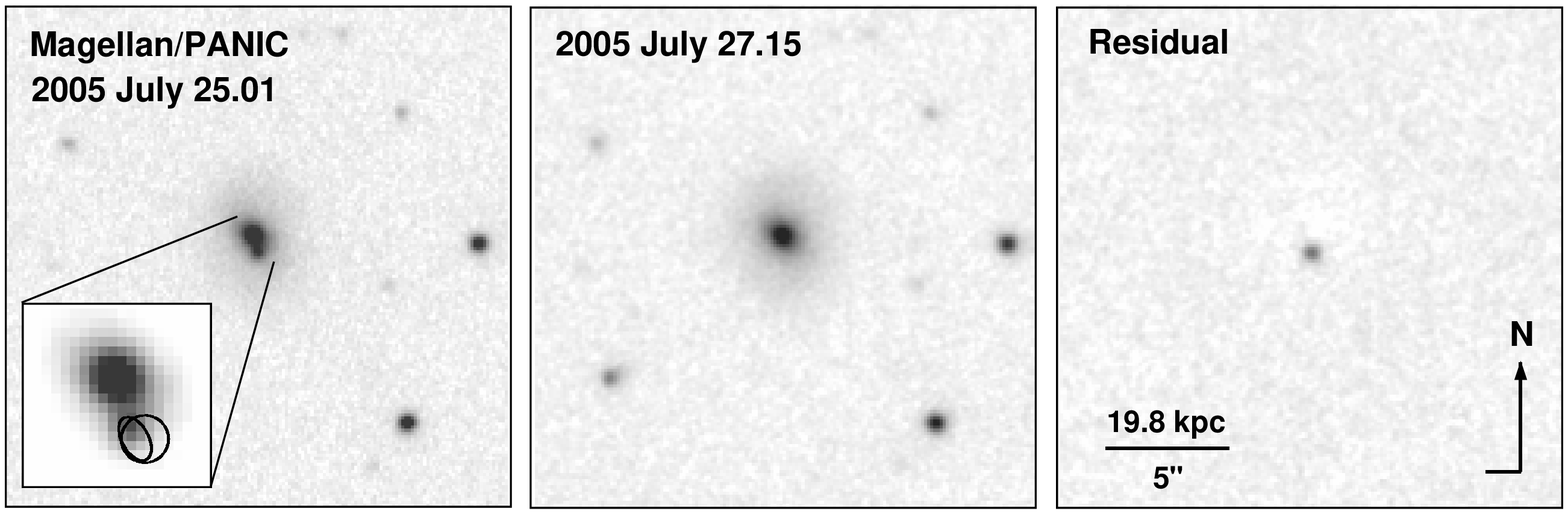}}
\caption{Discovery images of the near-infrared afterglow of
GRB\,050724 and its elliptical host galaxy.  The inset shows the Very
Large Array radio position (ellipse) and the {\it Chandra} X-ray
position (circle).  This was the first short burst to unambiguously
establish a link with an old stellar population.  From
\citet{bpc+05}.}
\label{fig:050724}
\end{figure}

It was only 15 days later that the discovery of X-ray, optical, and
for the first time radio afterglow emission finally established a
direct link between a short GRB and an old stellar population
\citep{bpc+05,bcb+05,gcg+06}.  The afterglow of GRB\,050724 was
localized to an elliptical galaxy at $z=0.257$ with no evidence for
star formation activity ($\lesssim 0.05$ M$_\odot$ yr$^{-1}$) and with
a stellar population age of $\gtrsim 1$ Gyr \citep{bpc+05,pbc+06}.
The absence of both star formation activity and an associated
supernova demonstrated a direct link to an old stellar population
\citep{bpc+05}.

\begin{figure}[ht!]
\centerline{\includegraphics[width=3.5in]{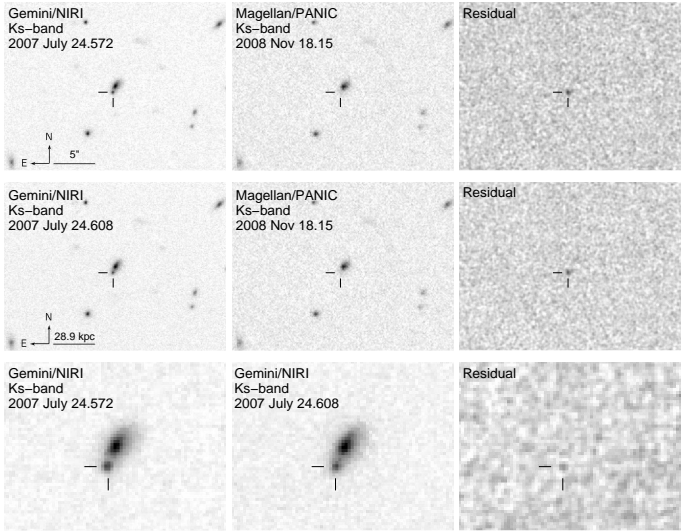}}
\caption{Discovery images of the near-infrared afterglow of
GRB\,070724 and its star forming host galaxy.  Each row shows an
afterglow image, a subsequent template image, and a residual image.
The fading afterglow coincides with the disk of the host galaxy.  From
\citet{bcf+09}.}
\label{fig:070724}
\end{figure}

The combination of low redshifts ($z\sim 0.2$) and the apparent
dominance of elliptical galaxies in the first few short GRB hosts led
to initial speculation of a particularly old progenitor population:
$\tau\gtrsim 4$ Gyr \citep{ngf06}, $\tau\gtrsim 7$ Gyr \citep{zr07},
and $\tau\gtrsim{\rm several}$ Gyr \citep{gno+08}.  Indeed, a possible
inconsistency with the expected merger time delay distribution of
NS-NS binaries was noted \citep{ngf06}, although subsequent population
synthesis models of NS-NS binary formation and mergers led to opposite
claims \citep{bpb+06}.  Clearly, the sample of short GRBs with
afterglow detections available when these various claims were
published was very small (GRBs 050509B, 050709, 050724, and 051221A).

\begin{figure}[ht!]
\centerline{\includegraphics[width=3.5in]{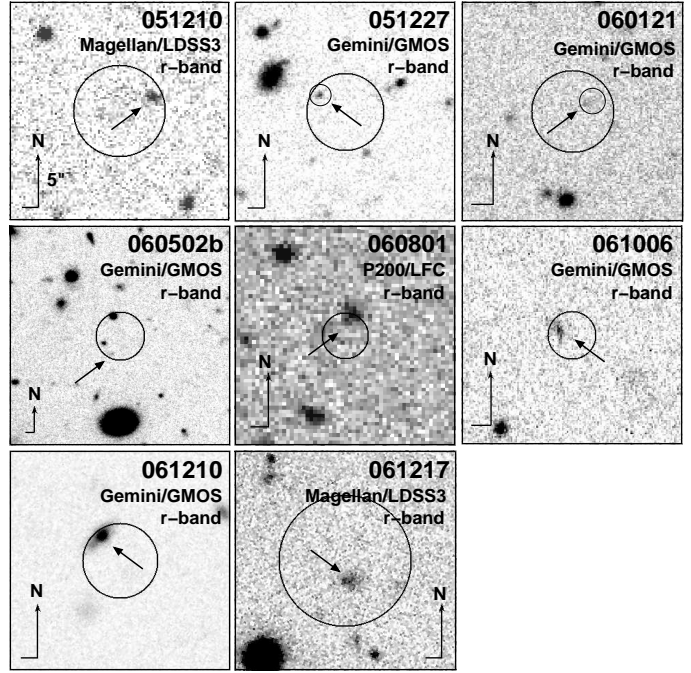}}
\caption{Ground-based images of several short GRB hosts obtained with
the Magellan 6.5-m telescopes and the Gemini 8-m telescopes.  All
images are $20''$ on a side, with the exception of GRB\,060502b which
is twice as large.  The large circles mark the XRT error regions,
while smaller circles mark the positions of the optical afterglows
(when available).  Arrows mark the positions of the identified hosts.
From \citet{bfp+07}.}
\label{fig:hosts}
\end{figure}

\begin{figure}[ht!]
\centerline{\includegraphics[width=3.5in]{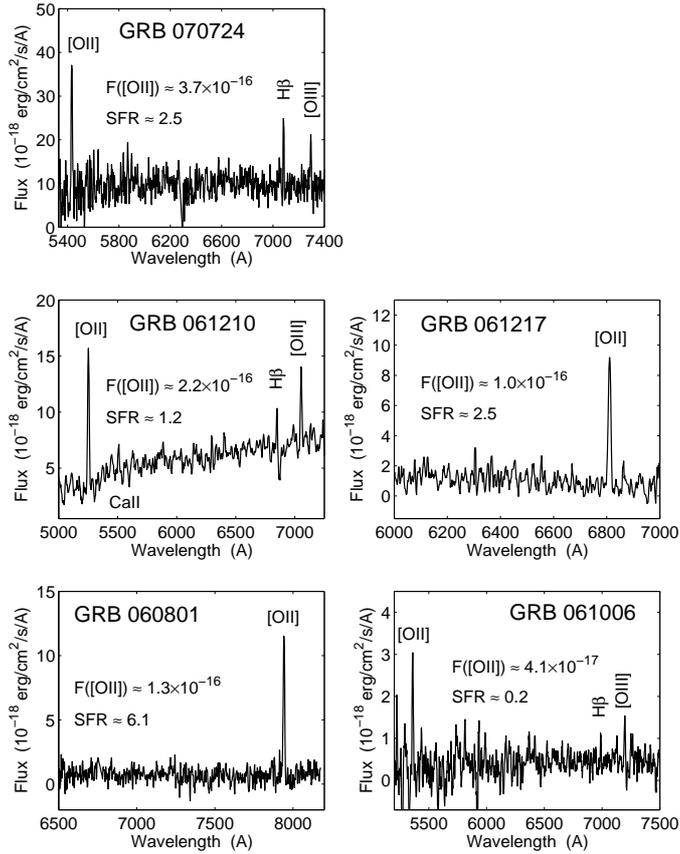}}
\caption{Optical spectra of several short GRB host galaxies.  The
relevant emission lines are marked and lead to redshifts of $z\approx
0.4-1.1$.  Also indicated are the star formation rates inferred from
the luminosity of the [OII]$\lambda 3727$ doublet.  From
\citet{ber09}.}
\label{fig:zs}
\end{figure}

\begin{figure}[ht!]
\centerline{\includegraphics[width=3.5in]{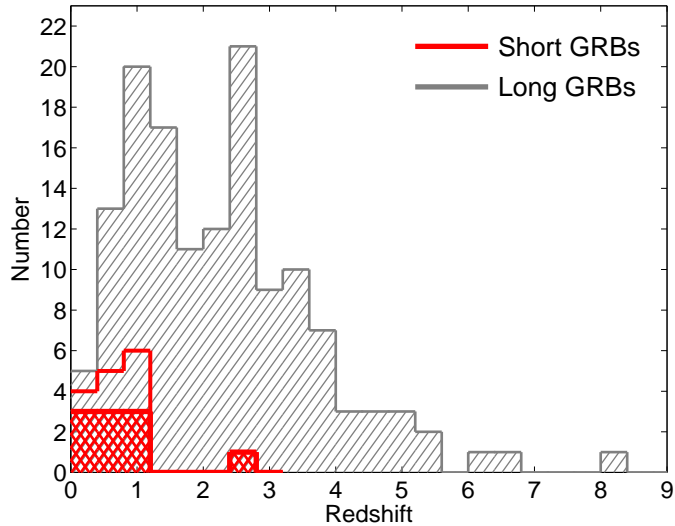}}
\caption{The redshift distributions of long (gray) and short (red)
GRBs as of late 2010.  The cross-hatched region indicates the
redshifts for short GRBs with sub-arcsecond positions, while the open
histogram includes the redshifts for host galaxies identified in some
XRT error circle ($\sim 2-5''$ radius).}
\label{fig:z}
\end{figure}

Fortunately, the continued detection of short GRBs (mainly, though not
exclusively by {\it Swift}), coupled with a community-wide concerted
effort to discover and study their afterglows, led to a substantial
increase in the sample of events over the past 5 years (e.g.,
Figures~\ref{fig:070724} and \ref{fig:hosts}).  Studies of this sample
have led to a re-evaluation of the host galaxy demographics and the
redshift distribution (e.g., \citep{bfp+07}).  In particular, as of
late 2010, the sample of short GRBs with X-ray detections (positions
of $\sim 2-5''$ radius) numbers about 40.  Of these, about 20 events
have been detected in the optical/UV/near-IR and/or radio, leading to
positional uncertainties of $\sim 0.1-0.5''$.  Host galaxies have been
identified for nearly all of the bursts with sub-arcsecond positions
(15/20), and putative hosts have also been identified for a
substantial fraction of the bursts with only X-ray positions (when
deep searches have been made).  At the present, 16 redshifts have been
measured between the two samples (Figures~\ref{fig:zs} and
\ref{fig:z}).

The events with only X-ray positions have two shortcomings.  First,
the probability of chance coincidence for the typical host magnitudes
within the XRT error circles ($\sim 21-26$ mag) is $\sim 10^{-3}-1$.
Second, in some cases there is disagreement about the position and
radius of the XRT error circles between various groups, leading to
systematic uncertainties in host associations.  Luckily, in the
subsequent discussion of detailed host properties no substantial
difference in the sample with and without optical afterglows is found,
suggesting that any spurious galaxy associations are at most a minor
contaminant.

\begin{figure}[ht!]
\centerline{\includegraphics[width=3.5in]{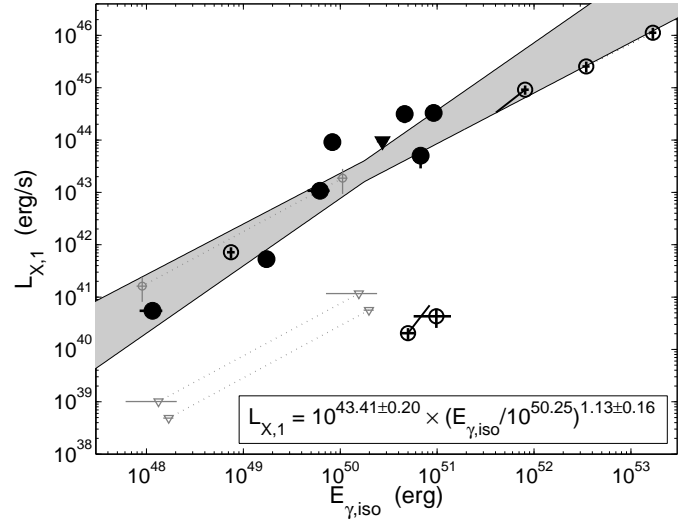}}
\caption{X-ray luminosity of short GRB afterglows normalized to $t=1$
day (a proxy for the afterglow kinetic energy, $\epsilon_eE_{K,{\rm
iso}}$) plotted as a function of $E_{\gamma,{\rm iso}}$ for events
with a known redshift (solid black circles), redshift constraints
(open black circles), and without any redshift information (gray
symbols connected by dotted lines for $z=0.1$ and 1, corresponding
roughly to the lowest and highest redshifts securely measured to
date).  The isotropic-equivalent relativistic energies are as least as
high as $10^{51}$ erg, and may approach $\sim 10^{53}$ erg for some
short bursts.  From \citet{ber07}.}
\label{fig:eglx}
\end{figure}

Using optical follow-up observations of nine short GRBs with X-ray
and/or optical afterglows (available by the end of 2006),
\citet{bfp+07} found that eight of the nine bursts were likely
associated with much fainter galaxies ($R\sim 23-26$ mag) than the
first few events.  By comparison to this early sample (with $R\sim
17-22$ mag and $z\lesssim 0.5$), as well as the hosts of long GRBs and
large field galaxy samples, it was demonstrated that these new host
galaxies likely reside at $z\sim 1$ and beyond.  A specific early case
for a $z\gtrsim 1$ origin was GRB\,060121 based on afterglow
photometric redshift estimates \citep{ucg+06,ltf+06}.  Spectroscopic
redshifts for the four {\it brightest} galaxies in this expanded
sample led to measurements of $z\approx 0.4-1.1$ \citep{bfp+07}; see
Figure~\ref{fig:zs}.  Subsequent observations have confirmed a broad
range of redshifts (e.g., \citep{gfl+09,adp+09,lbb+10}), and the
current redshift distribution (in comparison to that of long GRBs) is
shown in Figure~\ref{fig:z}.

One of the crucial ramifications of the measured redshift distribution
is the energy budget of the $\gamma$-ray emission and blastwave.  In
Figure~\ref{fig:eglx} I show the isotropic-equivalent afterglow X-ray
luminosities of short GRBs as a function of their isotropic-equivalent
$\gamma$-ray energies.  The former is a proxy for the afterglow
kinetic energy \citep{kum00,fw01,bkf03}.  I find that both quantities
span several orders of magnitude, with $E_{\gamma,{\rm iso}}\approx
10^{48}-10^{53}$ erg (although most short bursts have values of
$10^{50}-10^{51}$ erg).  This range is similar to that for long GRBs
\citep{fks+01,bkf03,bkp+03,bfk03}, and indicates that either short
GRBs can truly produce a broad range of energies, or instead exhibit a
wide range of collimation angles.  Due to the general faintness of
short GRB afterglows, strong evidence for collimation exists in only
one case \citep{sbk+06,bgc+06}, but additional cases are possible
\citep{ber07}.

\section{Host Galaxy Luminosities, Metallicities, and Star Formation 
Rates}
\label{sec:hosts}

The secure association of at least one short GRB (050724) with an
elliptical galaxy \citep{bpc+05,bcb+05,gcg+06} demonstrated
unambiguously that some of the progenitors are related to an old
stellar population.  However, as discussed in the previous section, a
substantial fraction of short GRBs ($1/3-2/3$) reside at higher
redshifts than initially suspected, $z\sim 1$ \citep{bfp+07}, and
spectroscopic observations indicate that most of these galaxies are
undergoing active star formation (Figure~\ref{fig:zs};
\citep{bfp+07,dmc+09,gfl+09}).  Indeed, in the sample of short GRBs
localized to better than a few arcseconds about $50\%$ of the bursts
occur in star forming galaxies compared to only $\approx 10\%$ in
elliptical galaxies; the remaining $\approx 40\%$ are currently
unclassified due to their faintness, a lack of obvious spectroscopic
features, or the absence of deep follow-up observations.  This result
raises the question of whether some short GRBs are related to star
formation activity rather than to an old stellar population, and if
so, whether the star formation properties are similar to those in long
GRB host galaxies.  The answer will shed light on the diversity of
short GRB progenitors.

\begin{figure}[ht!]
\centerline{\includegraphics[width=3.5in]{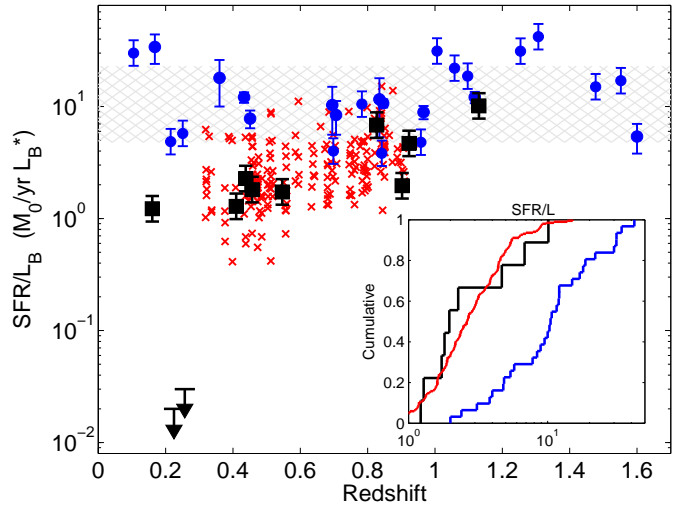}}
\caption{Specific star formation rates as a function of redshift for
the host galaxies of short GRBs (black squares), long GRBs (blue
circles) and field galaxies from the GOODS-N survey (red crosses;
\citep{kk04}).  Upper limits for the elliptical hosts of GRBs 050509B
and 050724 are also shown.  The cross-hatched region marks the median
and standard deviation for the long GRB host sample.  The inset shows
the cumulative distributions for the three samples.  The K-S
probability that the short and long GRB hosts are drawn from the same
distribution is only $0.3\%$, while the strong overlap with the field
sample leads to a K-S probability of $60\%$.  From \citet{ber09}.}
\label{fig:ssfr}
\end{figure}

In the following discussion, I compare several aspect of short and
long GRB host galaxies: Luminosities, metallicities, and star
formation rates.  A comparison of the masses and stellar population
ages is carried out in the subsequent section.  For the current sample
of short GRB hosts, the distribution of absolute rest-frame $B$-band
magnitudes ($M_B$) ranges from about $0.1$ to a few $L_*$
\citep{ber09,pbc+06}.  The star formation rates (mostly inferred from
the ${\rm [OII]}\lambda 3727$ line using the standard conversion
\citep{ken98}; Figure~\ref{fig:zs}) range from about $0.1$ to $10$
M$_\odot$ yr$^{-1}$ for the star forming hosts \citep{ber09,dmc+09}.
In the case of the elliptical hosts the upper limits are $\lesssim
0.1$ M$_\odot$ yr$^{-1}$ \citep{bpc+05,bpp+06,pbc+06,ber09}.  Combined
with the absolute magnitudes, the specific star formation rates (SSFR)
are ${\rm SFR}/L_B\approx 1-10$ M$_\odot$ yr$^{-1}$ $L_*^{-1}$ for the
star forming hosts, and $\lesssim 0.03$ M$_\odot$ yr$^{-1}$ $L_*^{-1}$
for the elliptical hosts.  The SSFR values as a function of redshift
are shown in Figure~\ref{fig:ssfr}.

For five\footnotemark\footnotetext{A sixth host, GRB\,071227, has an
inferred metallicity of about $0.4-1$ Z$_\odot$, but this was inferred
in the absence of detected hydrogen Balmer lines \citep{dmc+09}, and
the values are therefore prone to large systematic errors.} host
galaxies in the current sample, there is also sufficient spectral
information to measure the metallicity \cite{ber09,pbc+06,dmc+09}.  I
use the standard metallicity diagnostics, $R_{23}\equiv [(F_{\rm
[OII]\lambda 3727}+F_{\rm [OIII]\lambda\lambda 4959,5007})/F_{\rm
H\beta}$] \citep{peb+79,kk04} and $F_{\rm [NII]\lambda 6584}/F_{\rm
H\alpha}$.  The value of $R_{23}$ depends on both the metallicity and
ionization state of the gas, which is determined using the ratio of
oxygen lines, $O_{32}\equiv F_{\rm [OIII]\lambda\lambda
4959,5007}/F_{\rm [OII] \lambda 3727}$.  I note that the $R_{23}$
diagnostic is double-valued with low and high metallicity branches
(e.g., \citep{kd02}).  This degeneracy can be broken using the ratio
${\rm [NII]}/{\rm H}\alpha$ when these lines are accessible.  To
facilitate a subsequent comparison with field galaxy samples I use the
$R_{23}$, $O_{32}$, and ${\rm [NII]}/{\rm H}\alpha$ calibrations of
\citet{kk04}.  The typical uncertainty inherent in the calibrations is
about 0.1 dex.

\begin{figure}[ht!]
\centerline{\includegraphics[width=3.5in]{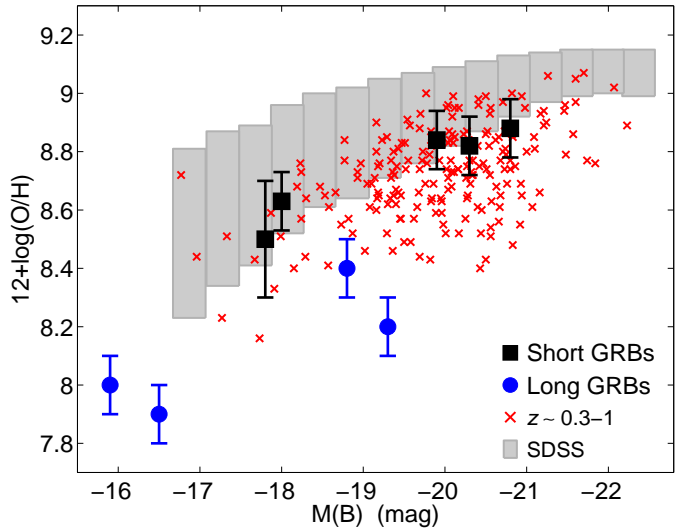}}
\caption{Metallicity as a function of $B$-band absolute magnitude for
the host galaxies of short GRBs (black squares) and long GRBs (blue
circles).  The gray bars mark the $14-86$ percentile range for
galaxies at $z\sim 0.1$ from the Sloan Digital Sky Survey
\citep{thk+04}, while red crosses designate the same field galaxies at
$z\sim 0.3-1$ shown in Figure~\ref{fig:ssfr} \citep{kk04}.  Both field
samples exhibit a clear luminosity-metallicity relation.  The long GRB
hosts tend to exhibit lower than expected metallicities
\citep{sgb+06}, while the hosts of short GRBs have higher
metallicities by about 0.6 dex, are moreover in excellent agreement
with the luminosity-metallicity relation.  From \citet{ber09}.}
\label{fig:lz}
\end{figure}

Adopting the solar metallicity from \citet{ags05}, $12+{\rm log(O/H)}=
8.66$ \citet{ber09} find $12+{\rm log(O/H)}\approx 8.6$ for the upper
$R_{23}$ branch and $\approx 8.0-8.5$ for the lower branch for the
host of GRB\,061006.  For the host of GRB\,070724 they find $12+{\rm
  log(O/H)} \approx 8.9$ for the upper branch, and $\approx 7.6-8.1$
for the lower branch.  A similar range of values is found for the host
of GRB\,061210, but the ratio $F_{\rm [NII]}/F_{\rm H\alpha}\approx
0.2$, indicates $12+{\rm log(O/H)}\gtrsim 8.6$, thereby breaking the
degeneracy and leading to the upper branch solution, $12+{\rm
  log(O/H)} \approx 8.9$.  For the host of GRB\,051221A
\citep{sbk+06}, similar values to those for the host of GRB\,070724
are inferred.  Finally, for the host galaxy of GRB\,050709, the ${\rm
  [NII]}/{\rm H}\alpha$ ratio indicates $12+{\rm log(O/H)}\approx
8.5$.  The dominant source of uncertainty in this measurement is the
unknown value of $O_{32}$, but using a spread of a full order of
magnitude results in a metallicity uncertainty of 0.2 dex.  For the
hosts with double-valued metallicities (GRBs 051221A, 061006, and
070724) I follow the conclusion for field galaxies of similar
luminosities and redshifts that the appropriate values are those for
the $R_{23}$ upper branch \citep{kk04}.  This conclusions was
advocated by \citet{kk04} based on galaxies in their sample with
measurements of both $R_{23}$ and ${\rm [NII]/H}\alpha$.  It is
similarly supported by our inference for the host galaxy of
GRB\,061210.  Future near-IR spectroscopy covering the [NII] and
H$\alpha$ lines will test this hypothesis.  The metallicities as a
function of host luminosity are shown in Figure~\ref{fig:lz}.

To place the host galaxies of short GRBs in a broader context I
compare their properties with those of long GRB hosts and field star
forming galaxies from the GOODS-N survey \citep{kk04}.  In terms of
absolute magnitudes, the long GRB hosts range from $M_B\approx -16$ to
$-22$ mag, with a median value of $\langle M_B \rangle\approx -19.2$
mag ($\langle L_B\rangle\approx 0.2$ $L_*$; \citep{bfk+07}).  Thus,
the long GRB hosts extend to lower luminosities than the short GRB
hosts, with a median value that is about 1.1 mag fainter.  A K-S test
indicates that the probability that the short and long GRB hosts are
drawn from the same underlying distribution is $0.1$.  On the other
hand, a comparison to the GOODS-N sample reveals a similar
distribution, and the K-S probability that the short GRB hosts are
drawn from the field sample is 0.6 \citep{ber09}.

A similar conclusion is reached based on a comparison of specific star
formation rates \citep{ber09}.  For long GRB hosts the inferred star
formation rates range from about 0.2 to 50 M$_\odot$ yr$^{-1}$, and
their specific star formation rates are about $3-40$ M$_\odot$
yr$^{-1}$ $L_*^{-1}$, with a median value of about 10 M$_\odot$
yr$^{-1}$ $L_*^{-1}$ \citep{chg04}.  As shown in
Figure~\ref{fig:ssfr}, the specific star formation rates of short GRB
hosts are systematically below those of long GRB hosts, with a median
value that is nearly an order of magnitude lower.  Indeed, the K-S
probability that the short and long GRB hosts are drawn from the same
underlying distribution is only $0.003$ \citep{ber09}.  This is
clearly seen from the cumulative distributions of specific star
formation rates for each sample (inset of Figure~\ref{fig:ssfr}).  On
the other hand, a comparison to the specific star formation rates of
the GOODS-N field galaxies reveals excellent agreement
(Figure~\ref{fig:ssfr}).  The K-S probability that the short GRB hosts
are drawn from the field galaxy distribution is 0.6.  Thus, short GRB
hosts appear to be drawn from the normal population of star forming
galaxies at $z\lesssim 1$, in contrast to long GRB hosts, which have
elevated specific star formation rates, likely as a result of
preferentially young starburst populations \citep{chg04,sgl08}.

Finally, the metallicities measured for short GRB hosts are in
excellent agreement with the luminosity-metallicity relation for field
galaxies at $z\sim 0.1-1$ (Figure~\ref{fig:lz}; \citep{kk04,thk+04}).
The two hosts with $M_B\approx -18$ mag have $12+{\rm log(O/H)}\approx
8.6$, while those with $M_B\approx -20$ to $-21$ mag have $12+{\rm
log(O/H)}\approx 8.8-8.9$, following the general trend.  On the other
hand, the short GRB host metallicities are systematically higher than
those of long GRB hosts, which have been argued to have lower than
expected values \citep{sgb+06}.  The median metallicity of short GRB
hosts is about 0.6 dex higher than for long GRB hosts, and there is
essentially no overlap between the two host populations \citep{ber09}.

To conclude, the short GRB host sample is dominated by star forming
galaxies, but these galaxies have higher luminosities, lower star
formation rates and specific star formation rates, and higher
metallicities than the star forming host galaxies of long GRBs.
Instead, the short GRB host sample appears to be drawn from the field
galaxy population.  These results suggest that while short GRB hosts
are mainly star forming galaxies, the progenitors most likely trace
stellar mass rather than the modest on-going star formation activity.

\section{Host Galaxy Stellar Masses and Ages}
\label{sec:hosts2}

\begin{figure*}[ht!]
\centering
\includegraphics[angle=0,width=2.2in]{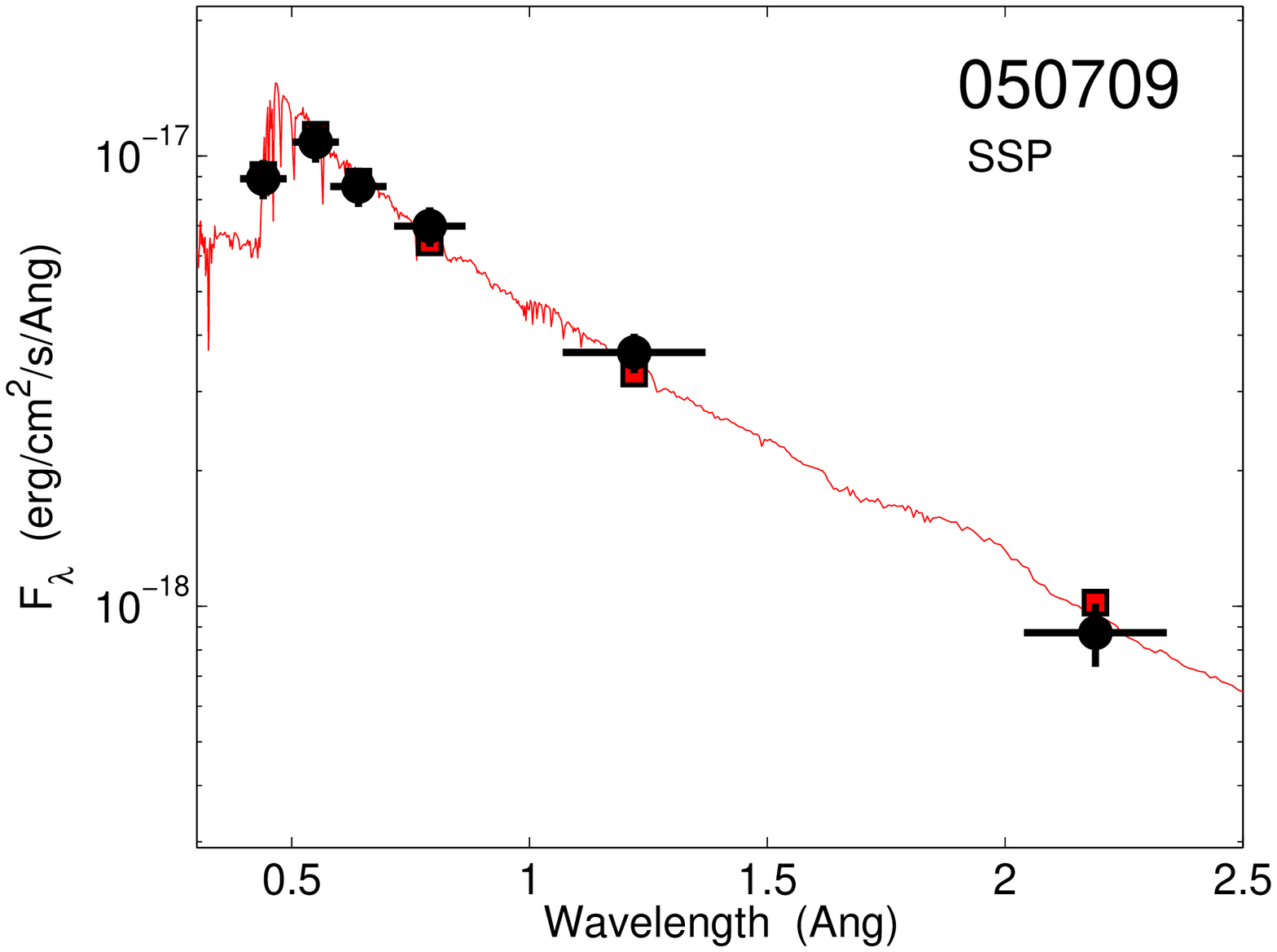}
\includegraphics[angle=0,width=2.2in]{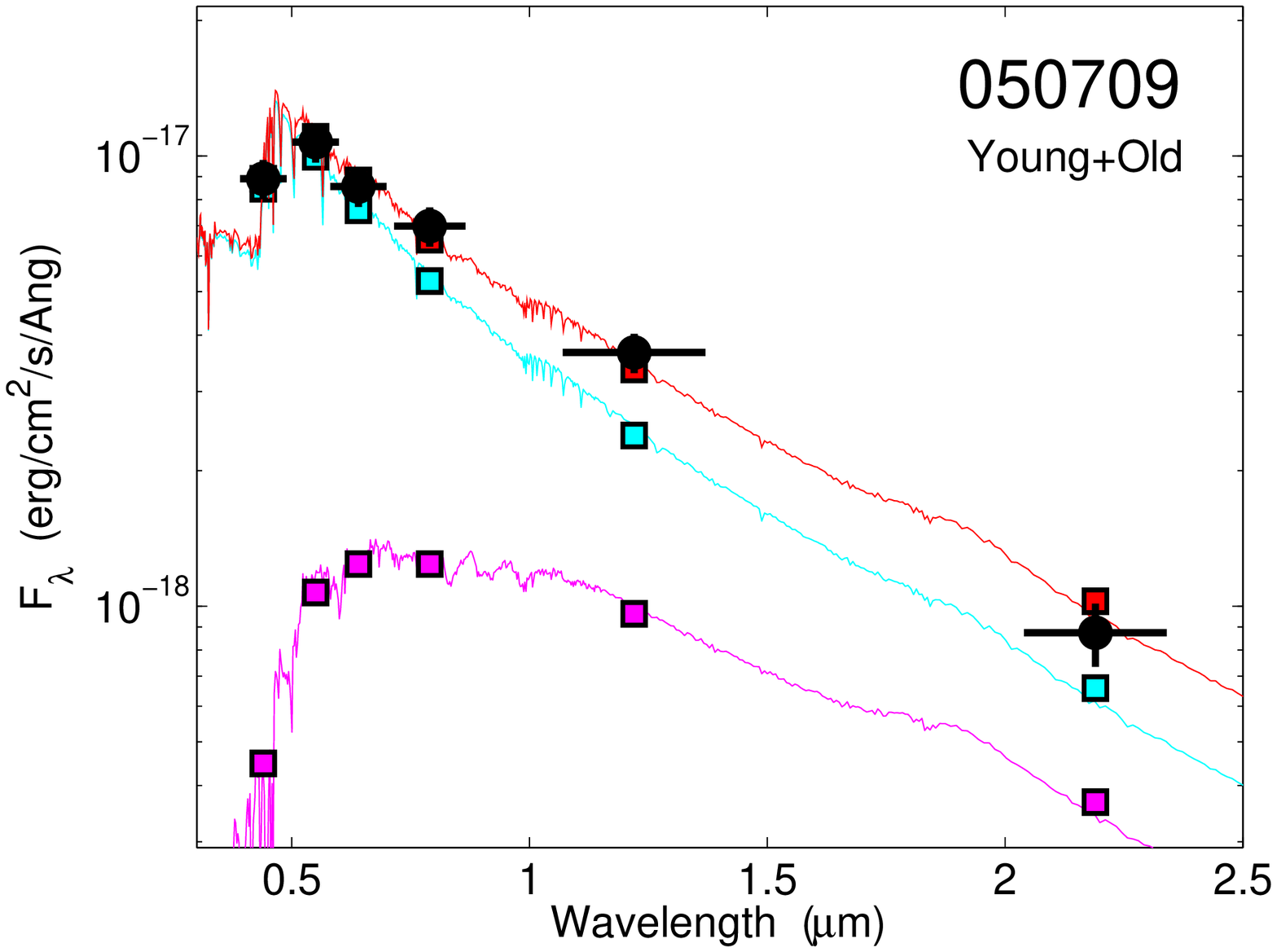}
\includegraphics[angle=0,width=2.2in]{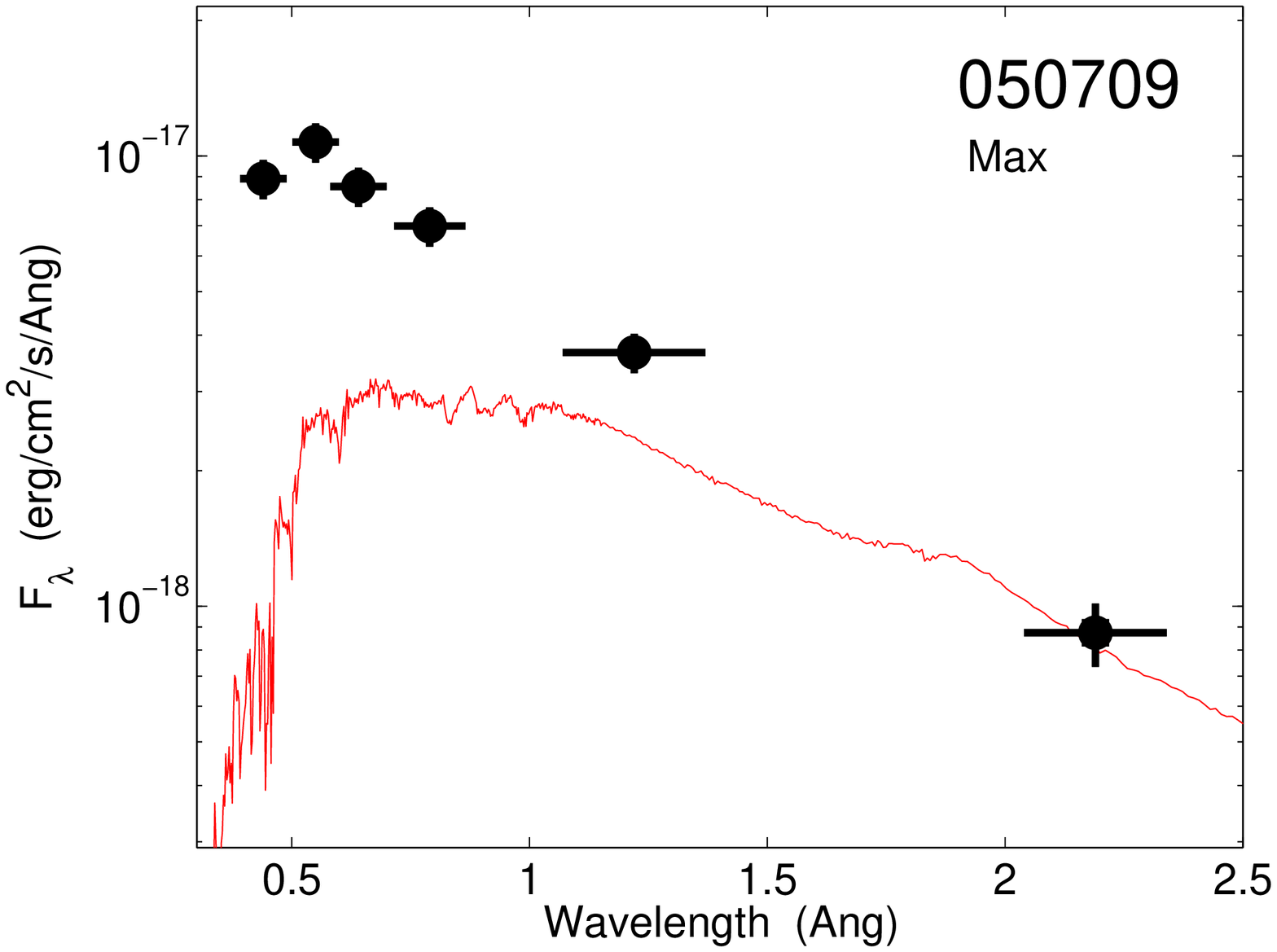}
\caption{Optical and near-IR spectral energy distribution of
GRB\,050709 with the three models used to extract the stellar mass and
population age.  Data are shown as black circles with error bars, and
synthesized model fluxes are shown as red squares.  {\it Left:} Single
age SSP model.  {\it Center:} Young+Old SSP model (magenta=old;
cyan=young) with the old population age fixed at the age of the
universe at the redshift of the burst ($z=0.161$ in this case).  {\it
Right:} Maximal mass model with the population age fixed at the age of
the universe and using only the $K$-band photometry.  The Young+Old
model leads to total masses intermediate between the single age SSP
and the maximal models, and has younger ages for the young population
than the single age SSP model.  From \citet{lb10}.
\label{fig:050709}} 
\end{figure*}

To more comprehensively address whether short GRBs trace stellar mass
alone (as would be expected for an old progenitor population), it is
essential to determine the stellar masses and population ages of short
GRB host galaxies, primarily in comparison to the general galaxy
stellar mass function.  This analysis was recently carried out by
\citet{lb10} using multi-band optical and near-IR data for 19 short
GRB hosts.  The resulting spectral energy distributions were fit with
the \citet{mar05} stellar population models to extract two crucial
parameters: stellar mass ($M_*$) and population age ($\tau$).  The
range of possible masses was assessed using three approaches.  First,
using single stellar population (SSP) fits, which provide an adequate
representation for the early-type hosts, but tend to under-estimate
the total mass and population age of star-forming hosts.  At the other
extreme, the near-IR data alone were modeled with a stellar population
matched to the age of the universe at each host redshift.  This
approach uses the maximum possible mass-to-light ratio to extract a
maximal mass for each host galaxy.  Finally, as a more realistic
estimate for the star-forming hosts, hybrid young+old stellar
populations were used.  Examples of all three approaches are shown for
the host of GRB\,050709 in Figure~\ref{fig:050709}.

\begin{figure}[ht!]
\centering
\includegraphics[angle=0,width=3.5in]{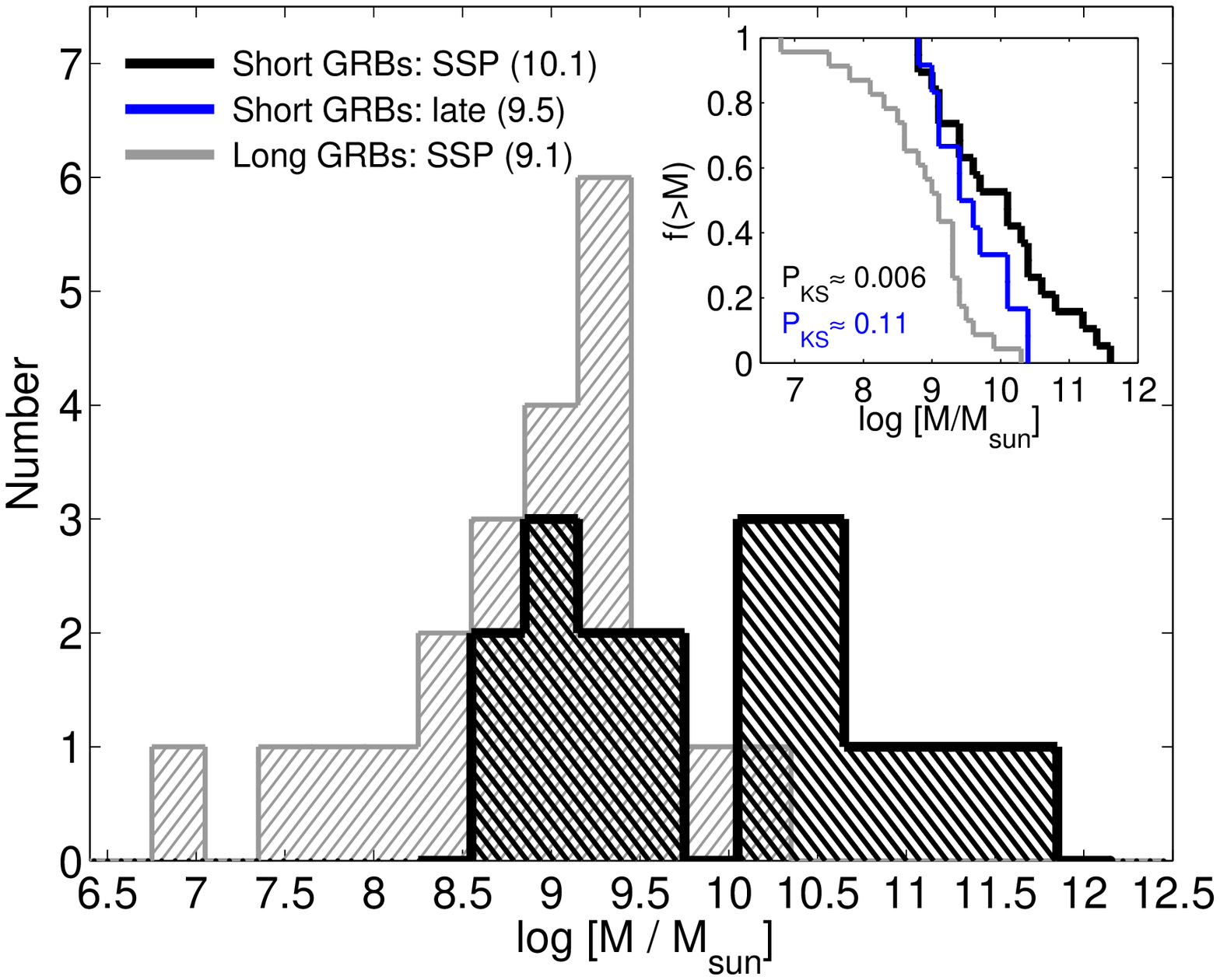}
\includegraphics[angle=0,width=3.5in]{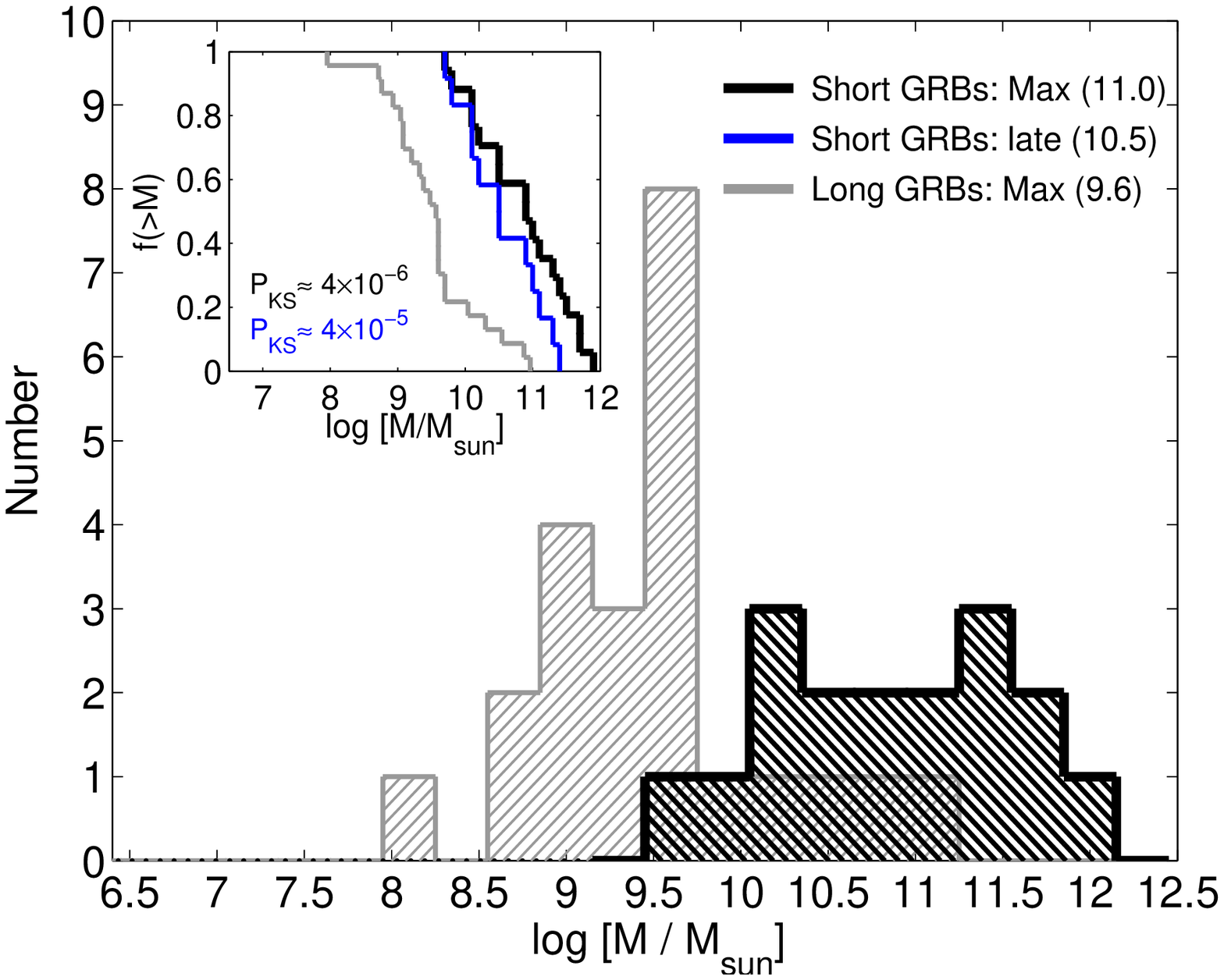}
\caption{{\it Top:} Histograms of inferred stellar masses from single
stellar population fits for the hosts of short (black) and long (gray)
GRBs.  The inset shows the cumulative distributions, including for the
subset of late-type short GRB hosts (blue).  The median values for the
three samples are given in parentheses, and the Kolmogorov-Smirnov
probabilities that the distributions of short and long GRB hosts, as
well as star forming short GRB and long GRB hosts are drawn from the
same distribution are provided in the inset.  {\it Bottom:} Same but
for the maximal masses.  From \citet{lb10}.
\label{fig:masses}} 
\end{figure}

The resulting mass distributions are shown in Figure~\ref{fig:masses}.
For comparison I also present the mass distributions for long GRB
hosts, which were analyzed with the same models for the purpose of a
uniform comparison.  The SSP masses span three orders of magnitude,
$M_{\rm SSP}\approx 6\times 10^8-4\times 10^{11}$ M$_\odot$, with a
median value of $\langle M_{\rm SSP} \rangle\approx 1.3\times 10^{10}$
M$_\odot$.  Dividing the sample into early- and late-type host
galaxies, the former span the range $M_{\rm SSP}\approx (2-40)\times
10^{10}$ M$_\odot$, while the latter have much lower masses of $M_{\rm
SSP}\approx (0.06-2)\times 10^{10}$ M$_\odot$.  The clear distinction
between the two samples partially reflects the bias of single age SSP
models, which for the late-type hosts are dominated by the young
stellar population and hence under-estimate the contribution of any
older stellar populations.  For the maximal masses the range is
$M_{\rm Max}\approx 6\times 10^9-8\times 10^{11}$ M$_\odot$.  The
median mass is $\langle M_{\rm Max}\rangle\approx 1\times 10^{11}$
M$_\odot$, about an order of magnitude larger than for the single age
SSP masses, and only slightly larger than the stellar mass of the
Milky Way.  As expected, the ratios of maximal to SSP masses for the
early-type hosts are modest, $M_{\rm Max}/M_{\rm SSP}\approx 2-8$,
since these hosts are already dominated by old stellar populations.
However, for the late-type hosts the corrections are significant,
$M_{\rm Max}/M_{\rm SSP}\approx 5-60$, with a median ratio of about an
order of magnitude.  Finally, using the young+old models for the
late-type hosts, and the single age SSP values for the early-type
hosts, the resulting masses are $M\approx 2\times 10^9-4\times
10^{11}$ M$_\odot$, with a median of $\langle M\rangle\approx 5\times
10^{10}$ M$_\odot$.

\begin{figure}[ht!]
\centerline{\includegraphics[width=3.5in]{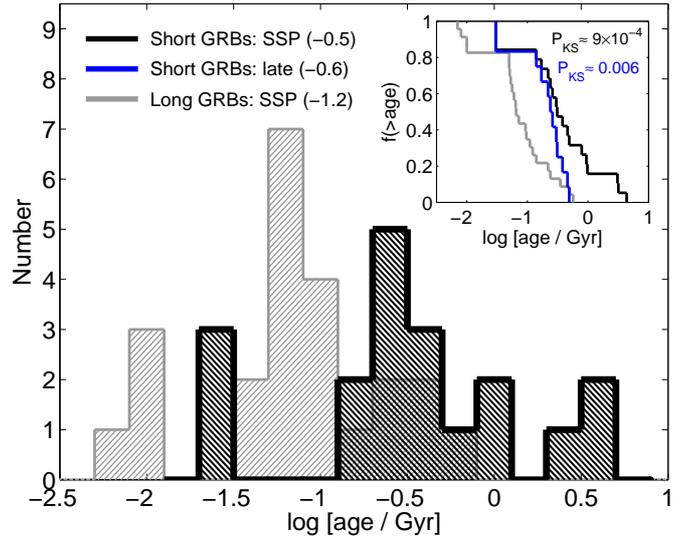}}
\caption{Histograms of inferred stellar population ages from single
stellar population fits for the hosts of short (black) and long (gray)
GRBs.  The inset shows the cumulative distributions, including for the
subset of late-type short GRB hosts (blue).  The median values for the
three samples are given in parentheses, and the Kolmogorov-Smirnov
probabilities that the distributions of short and long GRB hosts, as
well as star forming short GRB and long GRB hosts are drawn from the
same distribution are provided in the inset.  From \citet{lb10}.
\label{fig:ages}} 
\end{figure}

Stellar population ages can only be inferred for the single age SSP
models since for the maximal and young+old models the inherent
assumption is a population with the age of the universe at each host
redshift.  The distribution of ages is shown in Figure~\ref{fig:ages},
with the values ranging from about 30 Myr to 4.4 Gyr.  The median age
is $\langle\tau_{\rm SSP}\rangle\approx 0.3$ Gyr for the full sample,
with $\langle\tau_{\rm SSP}\rangle\approx 0.25$ Gyr for the subset of
late-type hosts and $\langle\tau_{\rm SSP}\rangle\approx 3$ Gyr for
the subset of early-type hosts.

The long GRB hosts, on the other hand, have lower masses and younger
population ages.  For the SSP model the mass range is $M_{\rm SSP}
\approx 6\times 10^6-2\times 10^{10}$ M$_\odot$, with a median value
of $\langle M_{\rm SSP}\rangle\approx 1.3\times 10^{9}$ M$_\odot$
(Figure~\ref{fig:masses}).  The maximal masses span $M_{\rm
Max}\approx 9\times 10^7-9\times 10^{10}$ M$_\odot$, with a median
value of $\langle M_{\rm Max}\rangle\approx 4.0\times 10^{9}$
M$_\odot$ (Figure~\ref{fig:masses}).  The SSP stellar population ages
span about 10 to 570 Myr, with a median value of $\langle \tau_{\rm
SSP}\rangle\approx 65$ Myr.

\subsection{Host Demographics}

In the redshift range relevant for short GRBs ($z\sim 0.2-1$) roughly
an equal fraction of the cosmic stellar mass budget resides in early-
and late-type galaxies (e.g., \citep{isl+10}).  Therefore, if short
GRBs track stellar mass alone we expect a roughly one-to-one ratio of
galaxy types.  This does not appear to be the case.  For example,
within the sample of short GRBs with optical afterglows (20 events),
only 2 are unambiguously hosted by early-type galaxies (GRBs 050724
and 100117; \citep{bpc+05}, Fong et al. in prep.), while 8 are
unambiguously hosted by late-type galaxies; the probability of
obtaining this ratio from an intrinsic one-to-one distribution is only
0.04.  The identity of the remaining 9 hosts is unclear at the present
due to their faintness or the lack of underlying galaxies at the burst
positions.  Still, unless nearly all of these bursts were hosted by
early-type galaxies, the resulting ratio appears to be skewed in favor
of late-type host galaxies with on-going star formation activity.  The
same result holds true even when considering the bursts with only
X-ray afterglow positions and identified hosts.

Thus, the host galaxy demographics suggest that short GRBs do not
track stellar mass alone, or phrased alternatively, they do not have a
delay time distribution that is skewed to old ages of $\sim {\rm few}$
Gyr.  It is possible, however, that this result is influenced by
secondary factors such as the typical circumburst density or intrinsic
differences in the energy scale and afterglow brightness as a function
of galaxy type (possibly reminiscent of the differences in peak
luminosity for Type Ia supernovae in early- and late-type galaxies;
\citep{htp+00,mdp06}).  If such differences lead to fainter afterglows
(or prompt emission) for short GRBs in early-type galaxies, this would
suppress the early-type fraction.  Although the modest size of the
host sample, and the substantial fraction of short GRBs with only
$\gamma$-ray positions ($\sim 1/3$ of all events), prevent definitive
conclusions, it does not appear that the optical afterglows of short
GRBs in early- and late-type galaxies are distinct \citep{ber10}; see
\S\ref{sec:kicks}.

\subsection{Comparison to the Galaxy Mass Function}
\label{sec:mass}

\begin{figure}[ht!]
\centerline{\includegraphics[width=3.7in]{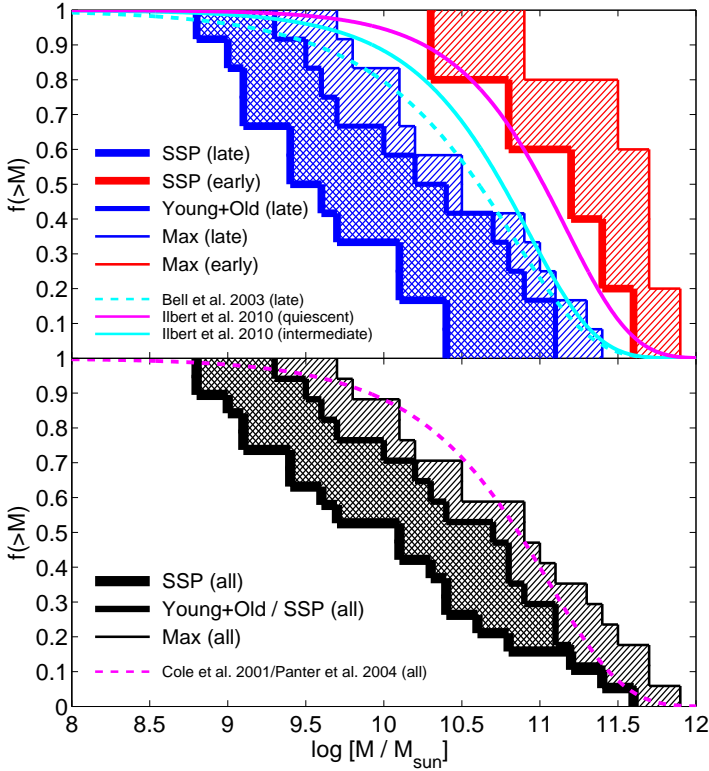}}
\caption{Cumulative distributions for the full sample of single age
simple stellar population (SSP) masses, maximal masses, and combined
young+old and SSP masses for the late- and early-type hosts,
respectively (black; bottom panel).  The upper panel shows a breakdown
by galaxy type (late-type: blue; early-type: red).  The shaded regions
represent the range of possible stellar masses since the SSP masses,
which are effectively light-weighted values, are most likely an
under-estimate, while the maximal masses make the extreme assumption
that all hosts are dominated by populations with the age of the
universe.  For the late-type hosts the total masses from a young+old
SSP fit are also shown; these are more closely representative of the
total mass.  Also shown are the fractions of total stellar mass in
galaxies with mass, $>M$, calculated from several published galaxy
stellar mass functions at $z\sim 0-2$ (cyan and magenta lines;
\citep{cnb+01,bmk+03,phj04,isl+10}); for the \citet{isl+10} mass
function the $z\sim 0.5$ bin is used, appropriate for the short GRB
sample, and separately plot the mass function for quiescent galaxies
and for intermediate-activity galaxies, which resemble the
intermediate star formation activity in short GRB hosts \citep{ber09}.
The comparison indicates that short GRBs trace galaxy mass {\it only}
if the bulk of the late-type hosts have close to maximal masses.  The
subset of early-type hosts appears to faithfully trace the mass
function of galaxies for the SSP-derived masses.  From \citet{lb10}.
\label{fig:masses_2}} 
\end{figure}

I next turn to a comparison of the inferred stellar masses with the
galaxy mass function \citep{lb10}.  The cumulative distribution of
stellar masses for the short GRB hosts is shown in
Figure~\ref{fig:masses_2} with the range of possible masses bounded by
the single age SSP and maximal values.  I also present a breakdown of
the sample into early- and late-type galaxies, each spanning the same
range.  For the late-type hosts, the intermediate young+old values are
also shown.  To compare these distributions to the distribution of
galaxy masses I also plot the cumulative stellar mass function {\it
weighted by mass}, i.e., the fraction of stellar mass in galaxies
above some mass, $f(>M)$, given by the equation:
\begin{equation}
f(>M)=\frac{\int_M^\infty{{M'\times\Phi(M')\,\,{\rm d}M'}}}
{\int_0^\infty{{M'\times\Phi(M')\,\,{\rm d}M'}}}
\end{equation}
where $\Phi(M)$ is the Schechter mass function:
\begin{equation}
\Phi(M)=\Phi^*\bigg(\frac{M}{M^*}\bigg)^\alpha{\rm
exp}\bigg(-\frac{M}{M^*}\bigg).
\end{equation} 

Several determinations of $\Phi(M)$ are used for this comparison,
including the \citet{cnb+01} mass function from the 2MASS/2dF catalogs
for all galaxy types at $z\sim 0$ ($M^*=10^{11.16}$ M$_\odot$,
$\alpha=-1.18$); the nearly identical \citet{phj04} mass function from
SDSS for all galaxy types at $z\sim 0$ ($M^*=10^{11.19}$ M$_\odot$,
$\alpha=-1.16$); the \citet{bmk+03} mass function for late-type
galaxies from 2MASS/SDSS converted to a Salpeter IMF for comparison
with our inferred values ($M^*=10^{10.97}$ M$_\odot$, $\alpha=-1.27$);
and the \citet{isl+10} mass functions from the COSMOS survey for
quiescent galaxies at $z\sim 0.3$ ($M^*=10^{11.13}$ M$_\odot$,
$\alpha=-0.91$) and intermediate-activity galaxies at $z\sim 0.5$
($M^*=10^{10.93}$ M$_\odot$, $\alpha=-1.02$), matched to the redshifts
of the early- and late-type short GRB host galaxies in our sample.
The resulting distributions of $f(>M)$ for the various mass functions
are shown in Figure~\ref{fig:masses_2}.

The agreement (or lack thereof) between the short GRB hosts and the
galaxy mass functions is assessed using the Kolmogorov-Smirnov (K-S)
test.  For the full sample there is negligible probability that the
distribution of single age SSP masses is drawn from the galaxy mass
function, with $P\approx 8\times 10^{-5}$.  On the other hand, for the
maximal mass distribution the probability is $P\approx 0.6$ indicating
that for these masses the short GRB sample is fully consistent with
the galaxy mass function.  Using the intermediate case of SSP masses
for the early-type hosts and the young+old masses for the late-type
hosts, the probability is $P\approx 0.3$, indicating that this
combination is also consistent with the galaxy mass function.

Separating the early-type hosts, their SSP masses are consistent with
the \citet{isl+10} mass function of quiescent galaxies ($P\approx
0.8$); their large maximal masses, on the other hand, are inconsistent
with the mass function, with $P\approx 0.007$.  Finally, for the
late-type hosts there is a clear inconsistency of the single age SSP
masses with the \citet{isl+10} mass function of intermediate-activity
galaxies, with $P\approx 4\times 10^{-7}$.  However, the maximal mass
distribution is consistent with the mass function ($P\approx 0.3$),
while the young+old mass distribution is marginally consistent
($P\approx 0.1$).

To summarize, the distribution of short GRB host masses is compatible
with the overall mass distribution of galaxies only if their stellar
masses are given by the SSP masses for the early-type hosts and the
maximal or young+old masses for the late-type hosts.  Since the
opposite scenario (maximal masses for the early-type hosts and SSP
masses for the late-type hosts) is unlikely, the existing sample of
short GRB hosts is consistent with the galaxy mass function
\citep{lb10}.  Equivalently, this means that short GRBs may indeed
track stellar mass alone.  However, I caution that the host
demographics seem to be at odds with the expected equal fractions of
total stellar mass in early- and late-type galaxies, unless nearly all
of the unidentified hosts are early-type galaxies.  This, along with
the somewhat lower than expected masses of the late-type hosts, leaves
open the possibility that at least a subset of short GRB progenitors
track star formation activity rather than stellar mass.

\subsection{Comparison to Long GRB Hosts}

Despite the possibility that some short GRB progenitors may track star
formation activity, the inferred stellar masses and population ages of
short GRB hosts are generally distinct from those of long GRB hosts in
both the single age SSP and maximal models \citep{lb10}.  Most
importantly, this is true for the subset of late-type hosts.  In the
framework of the single age SSP model the K-S probability is only
$0.006$ that the long and short GRB hosts are drawn from the same mass
distribution.  The probability is higher for the subset of late-type
short GRB hosts, $P\approx 0.1$ (Figure~\ref{fig:masses}).  However,
since the SSP values represent the mass of only the young stellar
populations, they are mostly reflective of the star formation activity
rather than the total stellar mass.  Using instead the maximal masses,
the K-S probability that the long GRB hosts and late-type hosts of
short GRBs are drawn from the same sample is negligible, $P\approx
4\times 10^{-5}$ (Figure~\ref{fig:masses}), demonstrating that they
are distinct galaxy populations.  A similar conclusion is apparent
from a comparison of the single age SSP population ages.  The K-S
probability that the long GRB hosts and late-type hosts of short GRBs
are drawn from the same distribution is only $P\approx 0.006$.

Thus, the long GRB hosts have significantly lower stellar masses than
the subset of late-type short GRB hosts, and their young stellar
population are significantly younger.  Indeed, a comparison of the
long GRB host maximal masses to the \citet{isl+10} mass function of
high-activity galaxies at $z\sim 0.7$ (appropriate for the long GRB
sample considered here) indicates a K-S probability of only $0.002$
that the long GRB hosts are drawn from the galaxy mass function.  This
is consistent with our understanding that their massive star
progenitors select galaxies by star formation (and perhaps additional
factors such as metallicity), but not by stellar mass.

The apparent distinction between long GRB hosts and the late-type
hosts of short GRBs in terms of their stellar masses and young
population ages strengthens the conclusion in \S\ref{sec:hosts} based
on the star formation rates, specific star formation rates,
luminosities, and metallicities \citep{ber09}.  In essentially every
property the late-type short GRB hosts point to a population of more
quiescent, massive, and evolved galaxies than the hosts of long GRBs.
I therefore conclude that this rules out the idea that short GRB
progenitors in late-type hosts are massive stars identical to long GRB
progenitors \citep{vzo+09}, {\it even if the short GRBs in late-type
galaxies indeed track star formation rather than stellar mass.}

\subsection{The Delay Time Distribution}
\label{sec:delay}

A determination of the delay function from the derived stellar
population ages is complicated by two primary factors.  First, they
rely on the assumption that the short GRB progenitors in each host
were formed within the inferred stellar population.  This assumption
is justified statistically both for an association of the progenitors
with stellar mass and with star formation activity, as long as we can
appropriately normalize the rates of short GRBs.  Second, while it is
possible to determine single age SSP ages from the broad-band
photometry, these data are not sufficient to provide an age breakdown
(by mass) for multiple stellar components.  Indeed, for the hybrid
young+old model, the age of the old population has to be fixed (to the
age of the universe in the analysis of \citet{lb10}).  Still, in the
young+old model, the bulk of the mass ($\approx 55-99\%$) is contained
in the old stellar population, and so the progenitors would have
``old'' ages ($\tau\gtrsim\tau_{\rm SSP}$) if they tracked stellar
mass.

As a result of these limitations it is only possible to explore the
implications of two main scenarios, namely that short GRBs track mass
and/or star formation activity.  In the context of the former scenario
I have shown in \S\ref{sec:mass} that the short GRBs in early-type
hosts trace stellar mass.  Therefore, their SSP ages can be used to
provide a rough estimate of the progenitor ages, $\tau\approx 0.8-4.4$
Gyr, with a median of about 3 Gyr.  On the other hand, for the
late-type hosts (for which no credible information on the
mass-weighted stellar population age can be extracted), it is possible
to infer a typical delay relative to the most recent star formation
episode under the assumption that these progenitors track star
formation activity.  I find SSP ages of $\tau\approx 0.03-0.5$ Gyr,
with a median of about 0.25 Gyr, or young+old ages of about $0.01-0.4$
Gyr with a median of about 0.1 Gyr.

Thus, if short GRBs follow both stellar mass (in early-type galaxies)
and star formation activity (in late-type galaxies), the typical delay
times are about 3 and 0.2 Gyr, respectively.

\section{Offsets and the Sub-Galactic Environments}
\label{sec:hst}

I next turn from a galactic-scale investigation of short GRB host
environments to the sub-galactic scale.  In general, the sub-galactic
environments of cosmic explosions provide powerful insight into the
nature of their progenitors.  For example, in the case of long GRBs,
the distribution of projected physical and host-normalized offsets
relative to the host centers matched the expected distribution for
massive stars in an exponential disk \citep{bkd02}.  Moreover, long
GRBs are spatially correlated with the brightest UV regions of their
hosts \citep{fls+06}.  Both of these studies have relied on high
angular resolution {\it Hubble Space Telescope} ({\it HST})
observations to spatially resolve the hosts and astrometrically locate
the GRB positions to pixel-scale accuracy.  As a by-product, these
observations also provided detailed morphological information for the
hosts (e.g., S\'{e}rsic index, effective radius; e.g. \citep{wbp07}).

\begin{figure}[ht!]
\centerline{\includegraphics[width=3.5in]{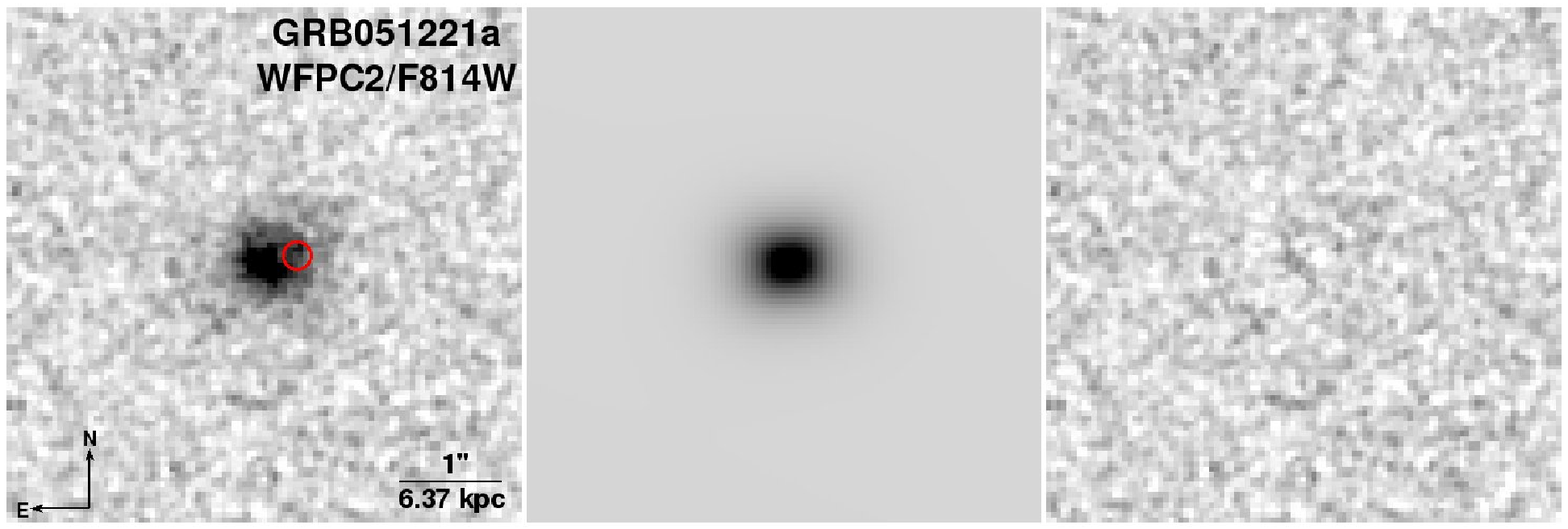}}
\centerline{\includegraphics[width=3.5in]{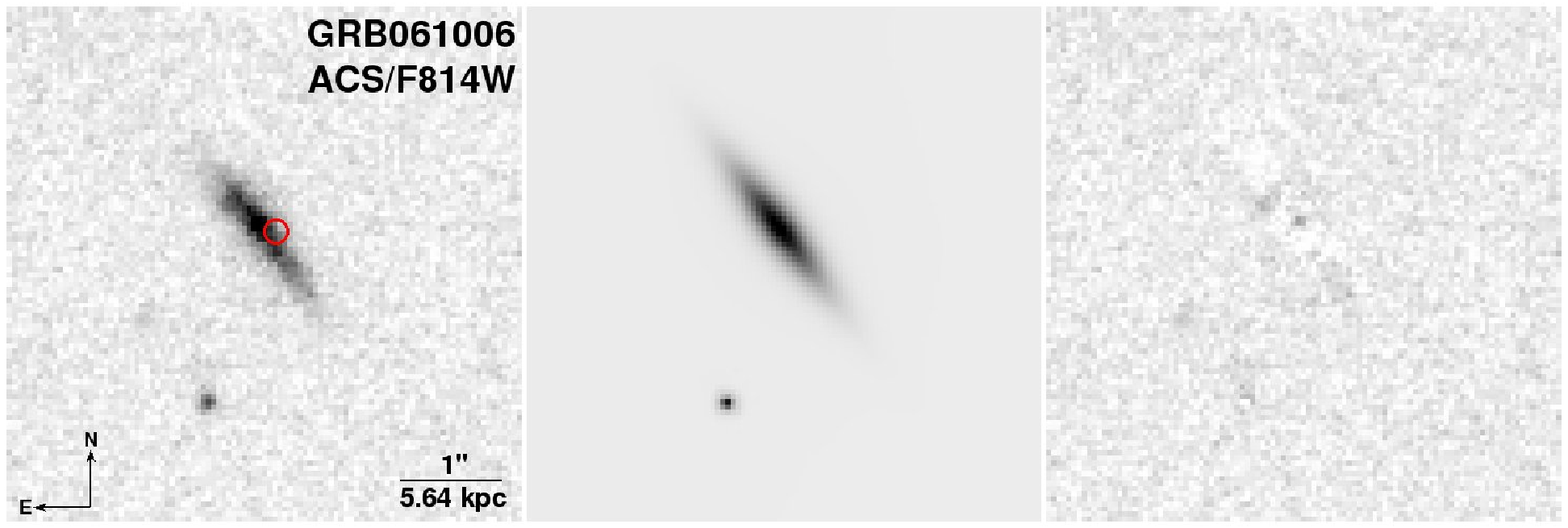}}
\caption{{\it Top-left:} {\it HST}/WFPC2/F814W image of the host
galaxy of GRB\,051221A with a $5\sigma$ error circle representing the
afterglow position.  {\it Top-center:} S\'{e}rsic model fit from {\tt
galfit}.  {\it Top-right:} Residual image.  {\it Bottom:} Same, but
for the host galaxy of GRB\,061006.  From \citet{fbf10}.}
\label{fig:051221}
\end{figure}

Individual offsets have been measured for several short GRBs (e.g.,
\citep{bpc+05,ffp+05,bpp+06}), and a study relying on ground based
data without a complete astrometric treatment was published by
\citet{tko+08}.  However, the first systematic study of short GRB
offsets, as well as their galactic environments and host morphologies
was recently published by \citet{fbf10}.  The sample includes ten
short GRBs (spanning May 2005 to December 2006), of which seven have
been localized to sub-arcsecond precision, and of those, six are
robustly associated with host galaxies (for details see
\citep{fbf10}).  Illustrative examples of {\it HST} host images and
morphological model fits are shown in Figure~\ref{fig:051221}.

\subsection{Morphological Analysis}
\label{sec:morph}

The two-dimensional surface brightness profiles were used to determine
the hosts' effective radii and morphological properties such as the
S\'{e}rsic $n$ index.  These quantities are crucial for a comparison
of the morphologies and sizes to those of long GRBs, as well as for
normalization of the projected offsets relative to the galaxy size.
Three hosts (GRBs 050709, 051221A, and 061006) are best modeled with
$n\approx 1$, corresponding to an exponential disk profile, while two
hosts (GRBs 050509B and 050724) are best modeled with $n\approx 3$ and
$\approx 5.6$, respectively, typical of elliptical galaxies.  These
classifications are in perfect agreement with their spectroscopic
properties.  The final three hosts in the sample (GRBs 051210, 060121,
and 060313) are faint, and as a results can be modeled with a wide
range of $n$ values, although $n\sim 1$ is preferred in all three
cases.  Therefore, of the eight short GRB hosts only two can be
robustly classified as elliptical galaxies based on their morphology.
A similar fraction was determined independently from spectroscopic
observations (\S\ref{sec:hosts}; \citep{ber09}).

\begin{figure}[ht!]
\centerline{\includegraphics[width=2.5in,angle=90]{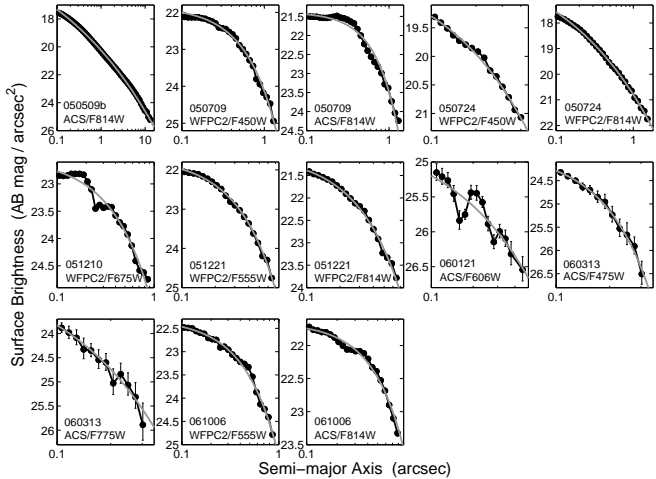}}
\caption{One-dimensional radial surface brightness profiles for short
GRB host galaxies derived using the IRAF task {\tt ellipse}.  The gray
lines are S\'{e}rsic model fits to the surface brightness profiles.
From \citet{fbf10}.}
\label{fig:sball}
\end{figure}

\begin{figure}[ht!]
\centerline{\includegraphics[width=3.5in]{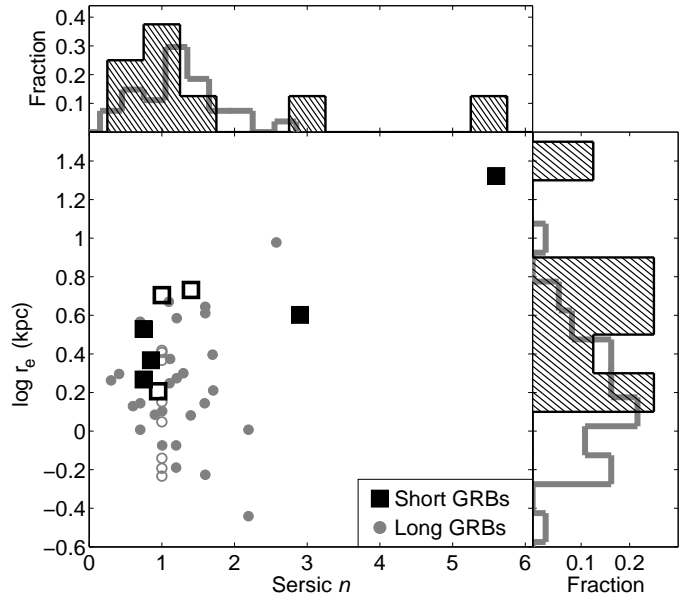}}
\caption{Effective radii for the short GRB hosts observed with {\it
HST} as a function of their S\'{e}rsic $n$ values
(Figure~\ref{fig:sball}).  Also shown are the data for long GRB hosts
based on {\it HST} observations from the sample of \citet{wbp07}.  The
hosts of short GRBs 050509B and 050724 have $n$ values typical of
elliptical galaxies, but the remaining hosts have a similar
distribution to that of long GRBs (i.e., a median of $n\sim 1$, or an
exponential disk profile).  On the other hand, the hosts of short GRBs
are larger by about a factor of 2 than the hosts of long GRBs, in
agreement with their higher luminosities.  From \citet{fbf10}.}
\label{fig:ren}
\end{figure}

The morphological analysis also yields values of the galaxy effective
radii, $r_e$.  A range of $\approx 0.2-5.8''$ is found, corresponding
to physical scales\footnotemark\footnotetext{For the faint hosts
without a known redshift (GRBs 051210, 060121, 060313, and possibly
061201) it is assumed that $z=1$ \citep{bfp+07}, and take advantage of
the relative flatness of the angular diameter distance as a function
of redshift beyond $z\sim 0.5$.} of about $1.4-21$ kpc.  The smallest
effective radius is measured for the host of GRB\,060313, while the
elliptical host of GRB\,050509B has the largest effective radius.  The
median value for the sample is $r_e\approx 3.5$ kpc.  The effective
radii as a function of $n$ are shown in Figure~\ref{fig:ren}.  Also
shown are the $r_e$ and $n$ values for the hosts of long GRBs from a
similar analysis carried out by \citet{wbp07}.

Two clear trends emerge from the morphological comparison of short and
long GRB hosts.  First, all long GRB hosts have $n\lesssim 2.5$, and
the median value for the population is $\langle n\rangle\approx 1.1$
\citep{wbp07}.  Thus, they are all morphologically classified as
exponential disk galaxies, while 2 of the 8 short GRB hosts exhibit de
Vaucouleurs elliptical galaxy profiles.  However, for the short GRB
hosts with $n\lesssim 2$, the distribution of $n$ values appears to be
similar to that of long GRB hosts \citep{fbf10}.

Second, the short GRB host galaxies have larger effective radii, with
$\langle r_e\rangle\approx 3.5$ kpc, compared to $\langle
r_e\rangle\approx 1.7$ kpc for long GRB hosts \citep{wbp07}.  A
Kolmogorov-Smirnov (K-S) test indicates that the probability that the
short and long GRB hosts are drawn from the same underlying
distribution of host galaxy effective radii is only 0.04.  Thus, short
GRB host galaxies are systematically larger than long GRB hosts.  The
larger sizes are expected in the context of the well-known galaxy
size-luminosity relation (e.g., \citep{fre70}) and the higher
luminosity of short GRB hosts (\S\ref{sec:hosts}; \citep{ber09}).

An additional striking difference between the hosts of long and short
GRBs is the apparent dearth of interacting or irregular galaxies in
the short GRB sample.  Of the eight short GRB host galaxies with {\it
HST} observations only one exhibits an irregular morphology
(GRB\,050709) and none appear to be undergoing mergers.  In contrast,
the fraction of long GRB hosts with an irregular or merger/interaction
morphology is $\sim 30-60\%$ \citep{wbp07}.  The interpretation for
the high merger/interaction fraction in the long GRB sample is that
such galaxies represent sites of intense star formation activity
triggered by the merger/interaction process, and are therefore
suitable for the production of massive stars \citep{wbp07}.  The lack
of morphological merger signatures in the short GRB sample indicates
that if any of the hosts have undergone significant mergers in the
past, the delay time between the merger and the occurrence of a short
GRB is $\gtrsim 1$ Gyr (e.g., \citep{bh92}).

\subsection{The Offset Distribution}

The location of each short GRB relative to its host galaxy center and
its overall light distribution was determined through differential
astrometry using optical and near-IR images of the
afterglows\footnotemark\footnotetext{Optical afterglows have not been
detected for GRBs 050509B and 051210.} \citep{fbf10}.  With the
exception of GRB\,050709, whose afterglow is directly detected in {\it
HST} observations, ground-based afterglow images from Magellan,
Gemini, and the VLT were used.  The resulting positional uncertainties
include contributions from the ground-based to {\it HST} astrometric
tie ($\sigma_{\rm GB\rightarrow HST}\approx 10-30$ mas), the
positional uncertainty of the afterglow ($\sigma_{\rm
\theta,GRB}\approx 1-40$ mas for optical afterglows and $\approx
1.7-5.8$ arcsec for X-ray afterglows), and the uncertainty in the
centroid of the host galaxy ($\sigma_{\rm \theta,gal}\approx 1-20$
mas).  The resulting combined offset uncertainties for the short GRBs
with optical afterglows are $\lesssim 60$ mas, corresponding to
physical offset uncertainties of $\lesssim 0.5$ kpc; the best-measured
offsets have uncertainties at the level of tens of pc.  These offsets
also correspond to about 1 {\it HST} pixel.

Based on the resulting astrometric ties \citet{fbf10} find that the
projected offsets of short GRBs relative to their host centers range
from about $0.12$ to $17.7''$.  The corresponding projected physical
offsets are about $1-64$ kpc, with a median value of about $5$ kpc.
The largest offsets are measured for GRBs 050509B and 051210, but
those are based on {\it Swift}/XRT positions only, with statistical
uncertainties of 12 and 18 kpc, respectively.  considering only the
bursts with sub-arcsecond afterglow positions, the largest offset is
3.7 kpc (GRB\,050709), and the median offset for the 6 bursts is 2.2
kpc.  In the case of GRB\,061201 the host association remains
ambiguous, but even for the nearest detected galaxy the offset is
about 14.2 kpc.

\begin{figure}[ht!]
\centerline{\includegraphics[width=3.5in]{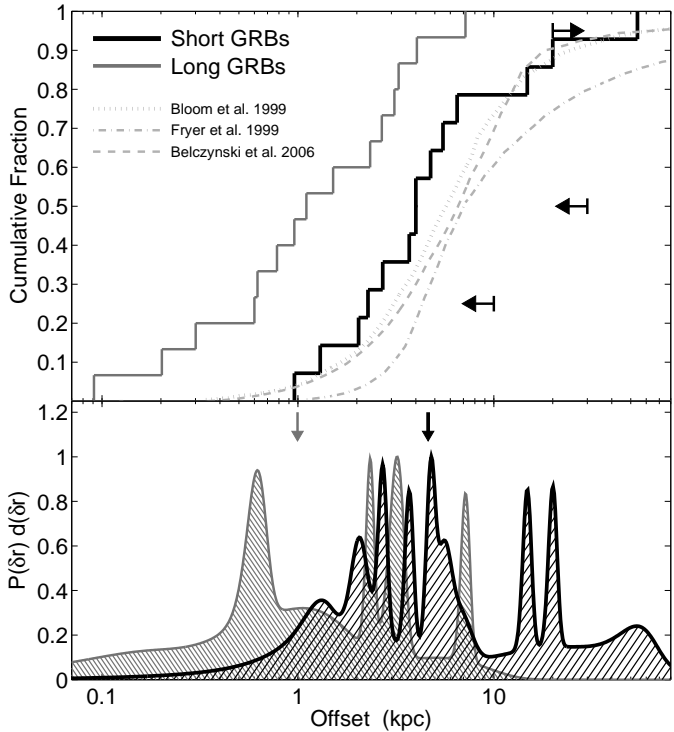}}
\caption{Projected physical offsets for short GRBs (black) and long
GRBs (gray; \citep{bkd02}).  The top panel shows a cumulative
distribution, while the bottom panel shows the differential
distribution taking into account the non-Gaussian errors on the
offsets.  The arrows in the bottom panel mark the median value for
each distribution.  The median value for short GRBs, $\approx 5$ kpc,
is about a factor of 5 times larger than for long GRBs.  The arrows in
the top panel mark the most robust constraints on the offset
distribution, taking into account the fraction of short GRBs with only
$\gamma$-ray positions, as well as short GRBs for which hosts have
been identified within XRT error circles (thereby providing a typical
range of $\sim 0-30$ kpc).  Also shown in the top panel are predicted
offset distributions for NS-NS binary mergers in Milky Way type
galaxies based on population synthesis models.  Good agreement between
the observed distribution and models, as well as between the robust
constraints and models is found.  From \citet{fbf10}.}
\label{fig:offset}
\end{figure}

To investigate the offset distribution in greater detail the {\it HST}
sample was supplemented with offsets for GRBs 070724, 071227, 080905A,
and 090510 from ground-based observations
\citep{bcf+09,gcn9353,rwl+10}.  These bursts have accurate positions
from optical afterglow detections.  In the case of GRBs 070724 and
071227 the optical afterglows coincide with the disks of edge-on
spiral galaxies (Figure~\ref{fig:070724}; \citep{bcf+09,dmc+09}).  The
offsets of the three bursts are 4.8, 14.8, 18.5, and 5.5 kpc,
respectively.

There are 7 additional events with optical afterglow identifications.
Of these, two bursts (070707 and 070714B) coincide with galaxies
\citep{pdc+08,gfl+09}, but their offsets have not been reported by the
authors.  Based on the claimed coincidence a conservative estimate is
$\lesssim 0.5''$, corresponding\footnotemark\footnotetext{GRB\,070714B
is located at $z=0.923$, while the redshift of GRB\,070707 is not
known.  Based on the faintness of the host, $R\approx 27.3$ mag, we
assume $z=1$ to calculate the physical offset.} to $\lesssim 4$ kpc.
For GRB\,090426 an offset relative to one of the knots in the host
galaxy complex was reported \citep{lbb+10}, but not relative to the
host center.  Finally, four bursts (070809, 080503, 090305, 090515) do
not have coincident host galaxies to deep limits; these bursts are
discussed in detail in \S\ref{sec:kicks}.

In addition to the bursts with sub-arcsecond positions, several hosts
have been identified within XRT error circles in follow-up
observations (GRBs 060801, 061210, 061217, 070429B, 070729, and
080123; \citep{bfp+07,ber09}).  Since the putative hosts are located
within the error circles, the offsets are consistent with zero or may
be as large as $\sim 30$ kpc (e.g., \citep{bfp+07}). For example, the
offsets for GRBs 060801, 061210, and 070429B are $19\pm 16$ kpc,
$11\pm 10$ kpc, and $40\pm 48$ kpc.  I adopt 30 kpc as a typical upper
limit on the offset for these 6 events.  No follow-up observations are
available in the literature for most short GRBs with X-ray positions
from 2008-2010.  Finally, about $1/4-1/3$ of all short GRBs discovered
to date have only been detected in $\gamma$-rays with positional
accuracies of a few arcminutes, thereby precluding a unique host
galaxy association and an offset measurement.

The cumulative distribution of projected physical offsets for the GRBs
with {\it HST} observations \citep{fbf10}, supplemented by the bursts
with offsets or limits based on optical afterglow positions (070707,
070714B, 070724, 070809, 071227, 080503, and 090510) is shown in
Figure~\ref{fig:offset}.  Also shown is the differential probability
distribution, $P(\delta r)d(\delta r)$, taking into account the
non-Gaussian errors on the radial offsets (see discussion in Appendix
B of \citep{bkd02}).  The median for this sample is about 5 kpc
\citep{fbf10}.

As evident from the preceding discussion, this is not a complete
offset distribution; roughly an equal number of short GRBs have only
limits or undetermined offsets due to their detection in just the
X-rays or $\gamma$-rays\footnotemark\footnotetext{I do not consider
the bursts that lack host searches since there is no a priori reason
that these events (mainly from 2008-2010) should have a different
offset distribution compared to the existing sample.}.  Taking these
events into account, our most robust inferences about the offset
distribution of short GRBs are as follows:
\begin{itemize}
\item At least $25\%$ of all short GRBs have projected physical
offsets of $\lesssim 10$ kpc.
\item At least $5\%$ of all short GRBs have projected physical offsets
of $\gtrsim 20$ kpc.  
\item At least $50\%$ of all short GRBs have projected physical
offsets of $\lesssim 30$ kpc; this value includes the upper limits for
the hosts identified within XRT error circles.
\end{itemize}
These robust constraints are marked in Figure~\ref{fig:offset}.

I next compare the short GRB offset distribution with the offsets of
long GRBs from the sample of \citet{bkd02}; see
Figure~\ref{fig:offset}.  The offset distribution of long GRBs has
been used to argue for a massive star progenitor population, and
against NS-NS binaries \citep{bkd02}.  The offset distribution of
short GRBs is clearly shifted to larger physical scales.  In
particular, the median offset for the long GRBs is 1.1 kpc, about a
factor of 5 times smaller than the median value for short GRBs.
Similarly, no long GRB offsets are larger than about 7 kpc, whereas at
least some short GRBs appear to have offsets in excess of 15 kpc.  The
significant difference between the offset distributions indicates that
short GRBs do not arise from the same progenitor population as long
GRBs.

I further compare the observed distribution (and the robust
constraints outlined above) with predicted distributions for NS-NS
binaries in Milky Way type galaxies \citep{bsp99,fwh99,bpb+06},
appropriate for the observed luminosities of short GRB host galaxies
\citep{ber09}.  There is good agreement between the observed
distribution and those predicted by \citet{bsp99} and \citet{bpb+06}.
The offset distribution of \citet{fwh99}, with a median of about 7
kpc, predicts larger offsets and therefore provides a poorer fit to
the observed distribution, which has a median of about 5 kpc.
However, all three predicted distributions accommodate the offset
constraints.  In particular, they predict about $60-75\%$ of the
offsets to be $\lesssim 10$ kpc, about $80-90\%$ to be $\lesssim 30$
kpc, and about $10-25\%$ of the offsets to be $\gtrsim 20$ kpc.  Thus,
the projected physical offsets of short GRBs are consistent with
population synthesis predictions for NS-NS binaries.  However, the
observations are also consistent with partial contribution from other
progenitor systems for which kicks are not expected (magnetars, WD-WD
binaries, accreting NS).

\subsection{Light Distribution Analysis}

In addition to the offset analysis in the previous section,
\citet{fbf10} studied the local environments of short GRBs using a
comparison of the host brightness at the GRB location to the overall
host light distribution.  This approach is advantageous because it is
independent of galaxy morphology, and does not suffer from ambiguity
in the definition of the host center (see \citep{fls+06}).  I note
that for the regular morphologies of most short GRB hosts
(\S\ref{sec:morph}), the definition of the host center is generally
robust, unlike in the case of long GRBs \citep{fls+06,wbp07}.  On the
other hand, this approach has the downside that it requires precise
pixel-scale positional accuracy.  In the existing sample, this is only
available for 6 short bursts.

\begin{figure}[ht!]
\centerline{\includegraphics[width=3.5in]{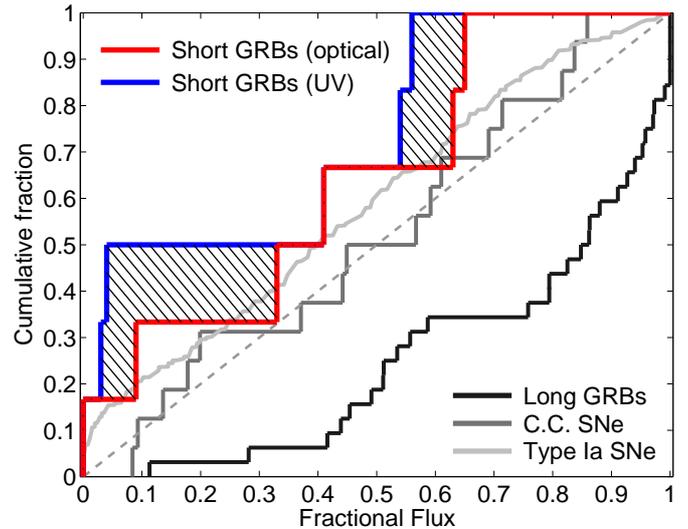}}
\caption{Cumulative distribution of fractional flux at the location of
short GRBs relative to their host light.  For each burst the fraction
of host light in pixels fainter than the GRB pixel location is
measured.  The shaded area is defined by the results for the two
available filters for each short GRB.  Also shown are data for long
GRBs (dark gray line) and for core-collapse and Type Ia SNe (light
gray lines) from \citet{fls+06} and \citet{kkp08}.  The dashed line
marks the expected distribution for objects which track their host
light distribution.  Short GRBs appear to under-represent their host
light, while long GRBs tend to be concentrated in the brightest
regions of their hosts \citep{fls+06}.  From \citet{fbf10}.}
\label{fig:light}
\end{figure}

The fraction of total host light in pixels fainter than the afterglow
pixel brightness for each host/filter combination is given in
\citet{fbf10}.  The cumulative light distribution histogram is shown
in Figure~\ref{fig:light}.  The shaded histogram represents the range
defined by the dual filters for 5 of the 6 bursts.  The upper bound of
the distribution is defined by the blue filters, indicating that short
GRBs trace the rest-frame optical light of their hosts better than the
rest-frame UV.  This indicates that short GRB progenitors are likely
to be associated with a relatively old stellar population, rather than
a young and UV bright population.

The overall distribution has a median value of $\approx 0.1-0.4$ (blue
vs.~red filters); namely, only in about one-quarter of the cases,
$50\%$ of the host light is in pixels fainter than at the GRB
location.  Thus, the overall distribution of short GRB locations
under-represents the host galaxies light distribution, but traces the
red light (old stars) more closely than the blue light (star
formation).  This is also true in comparison to the distribution for
core-collapse SNe, which appear to track their host light
\citep{fls+06}, and even Type Ia SNe, which have a median of about 0.4
\citep{kkp08}.  Thus, the progenitors of short GRBs appear to be more
diffusely distributed than Type Ia SN progenitors.

An extensive analysis of the brightness distribution at the location
of long GRBs has been carried out by \citet{fls+06}.  These authors
find that long GRBs are more concentrated on the brightest regions of
their hosts than expected from the light distribution of each host.
In particular, they conclude that the probability distribution of long
GRB positions is roughly proportional to the surface brightness
squared.  As can be seen from Figure~\ref{fig:light}, short GRBs have
a significantly more diffuse distribution relative to the host light
than long GRBs.  In particular, for the latter, the median light
fraction is about 0.85 compared to about $0.25\pm 0.15$ for the short
GRBs.

\section{Is There Evidence for Large Progenitor Kicks?}
\label{sec:kicks}

\begin{figure*}[ht!]
\centering
\includegraphics[angle=0,width=2.2in]{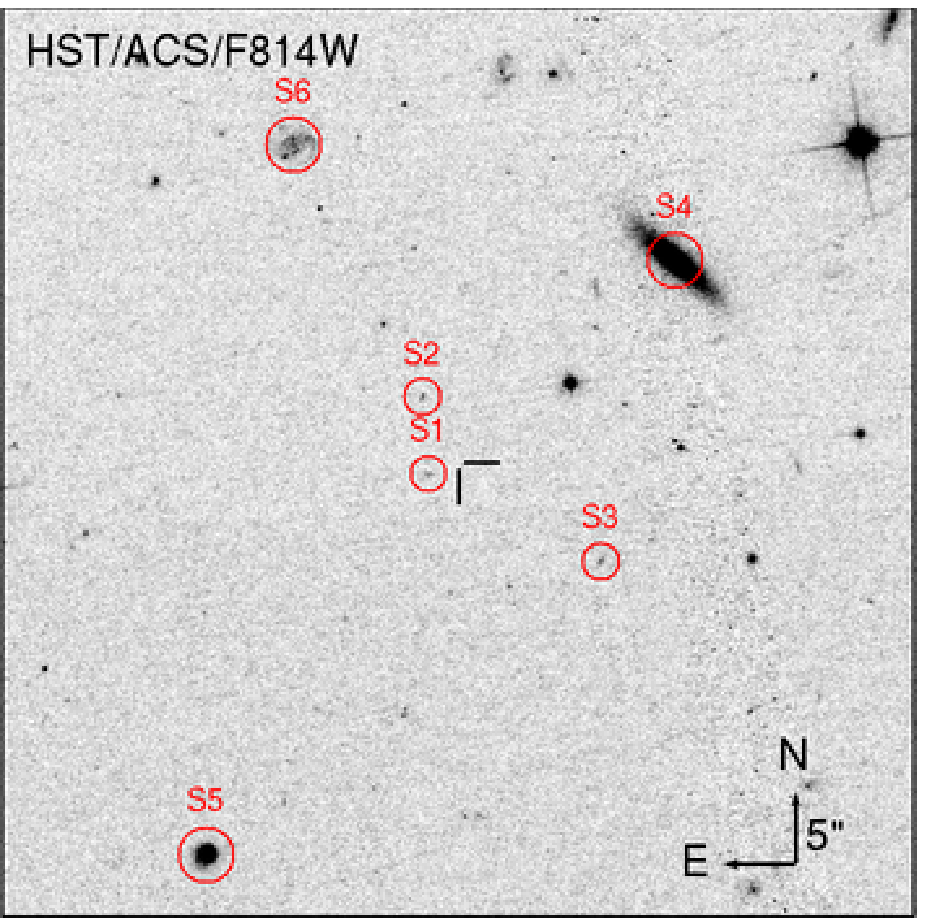}
\includegraphics[angle=0,width=2.2in]{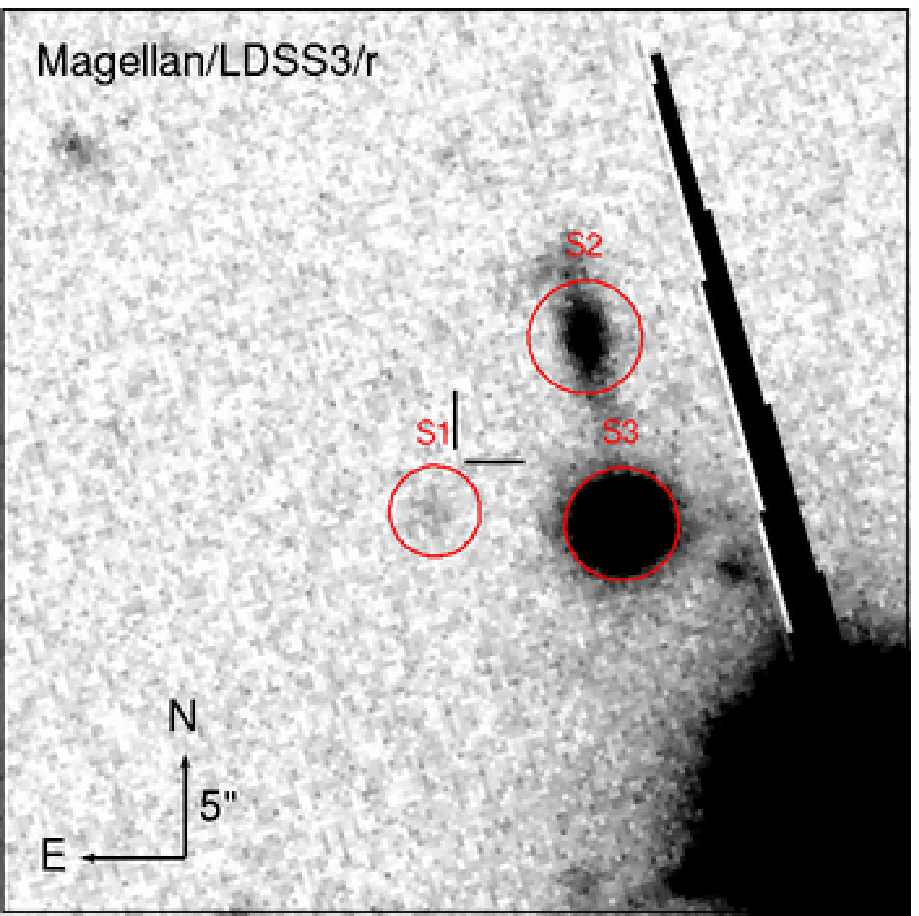}
\includegraphics[angle=0,width=2.2in]{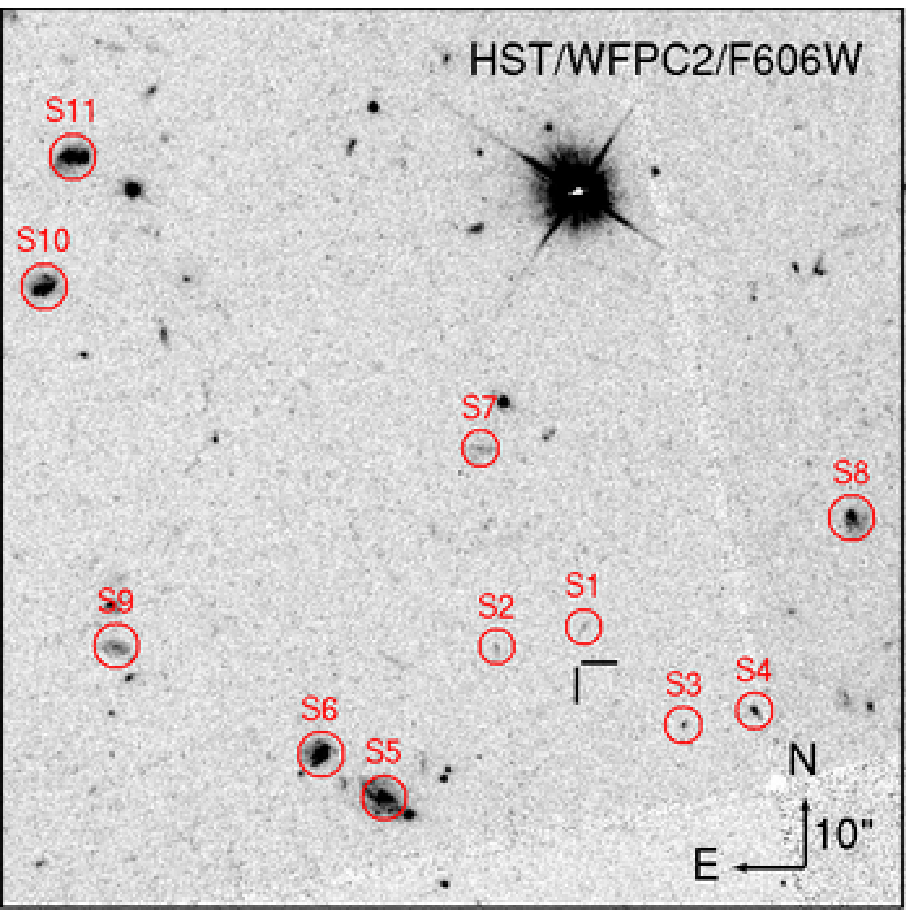}
\includegraphics[angle=0,width=2.2in]{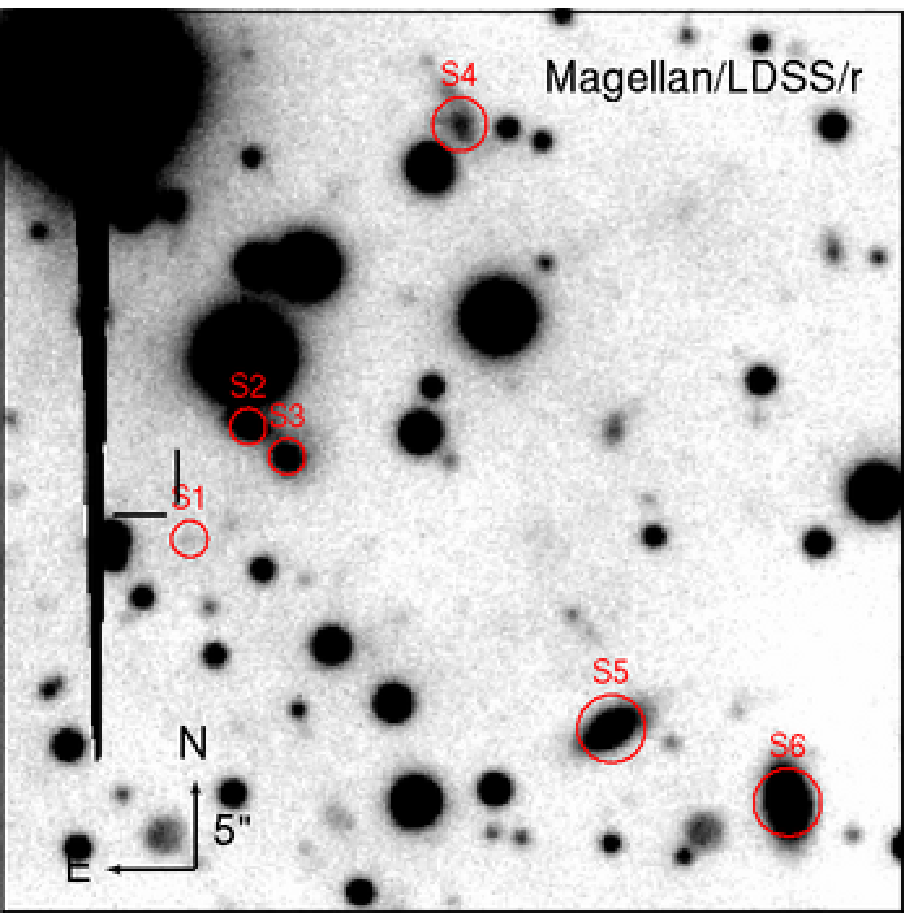}
\includegraphics[angle=0,width=2.2in]{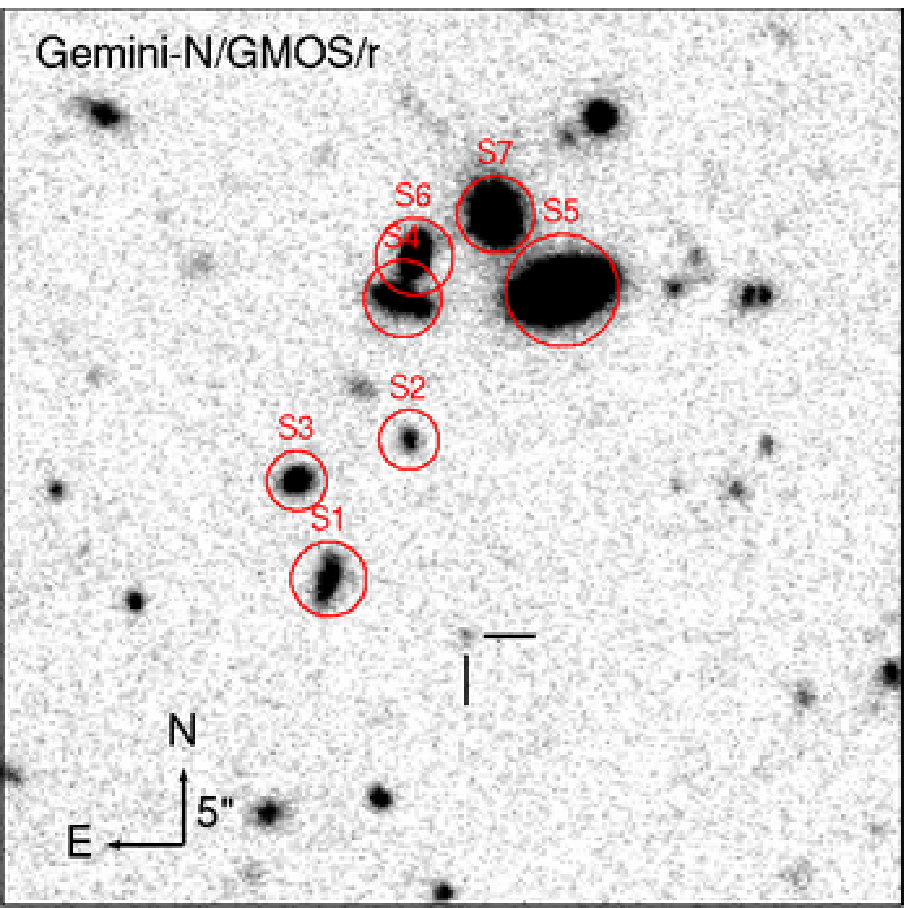}
\caption{Images of the fields around the optical afterglow positions
of short GRBs 061201, 070809, 080503, 090305A, and 090515 (top-left to
bottom-right).  These are the 5 bursts in the current sample that have
optical afterglow positions and no coincident hosts to limits of
$\gtrsim 25.5$ mag.  From \citet{ber10}.}
\label{fig:nohost}
\end{figure*}

One of the predictions of the compact object coalescence model is that
some binaries could merge at large separations from their host
galaxies due to a combination of large kick velocities and a long
merger timescale \citep{bsp99,fwh99,bpb+06}.  With kicks of several
hundred km s$^{-1}$ and merger timescales of $\sim {\rm Gyr}$, such a
binary could travel several hundred kpc from its host, corresponding
to $\sim 1'$ at $z\sim 1$ and $\sim 10'$ at $z\sim 0.1$.  Such large
offsets would not be expected in other progenitor scenarios.  It is
important to note, however, that if the typical kick velocities are
$\lesssim 10^2$ km s$^{-1}$, an NS-NS/NS-BH system is likely to remain
bound to its host regardless of the merger timescale, and hence to
reside at offsets of $\lesssim$\,tens of kpc (as already inferred for
some short GRBs \citep{fbf10}; Figure~\ref{fig:offset}).  Similarly,
short merger timescales (tens to hundreds of Myr) would also lead to
relatively small offsets regardlss of the kick velocity.

Clearly, an observational demonstration of a large offset is not
trivial.  Ideally, we would like to measure the redshift of the burst
directly through afterglow spectroscopy and then associate it with a
galaxy at a large separation.  However, to date, short GRB redshifts
have been measured through their host associations so this test is not
possible.  

At a more tentative level, we can also investigate the large-scale
environments around short GRBs that do not appear to spatially
coincide with bright galaxy counterparts to assess the potential for a
host with large offset.  This is a particularly important test if
combined with the afterglow properties of short GRBs with and without
coincident hosts.  In the current sample, there are 5 cases of short
GRBs with sub-arcsecond positions and no obvious bright
hosts\footnotemark\footnotetext{I do not discuss cases with only XRT
positions of a few arcsec radius since those do not generally lead to
significant offsets and furthermore nearly always contain at least one
possible host consistent with zero offset.}.  Below, I assess the
possibility of large offsets for these bursts, and compare this with
alternative explanations (e.g., a high redshift origin).  The sample
includes 20 short GRBs with optical afterglows.  Images of the fields
around the 5 bursts with no coincident hosts are shown in
Figure~\ref{fig:nohost}; hereafter, I denote these 5 bursts as {\it
Sample 2}, with the remaining 15 bursts with coincident hosts
designated as {\it Sample 1}.  The afterglow positions, as well as
nearby galaxies with varying probabilities of chance coincidence are
marked in Figure~\ref{fig:nohost}.  The limits at the positions of the
afterglows range from $\gtrsim 25.5$ to $\gtrsim 26.5$ mag
\citep{ber10}.

\begin{figure}[ht!]
\centering
\includegraphics[angle=0,width=3.5in]{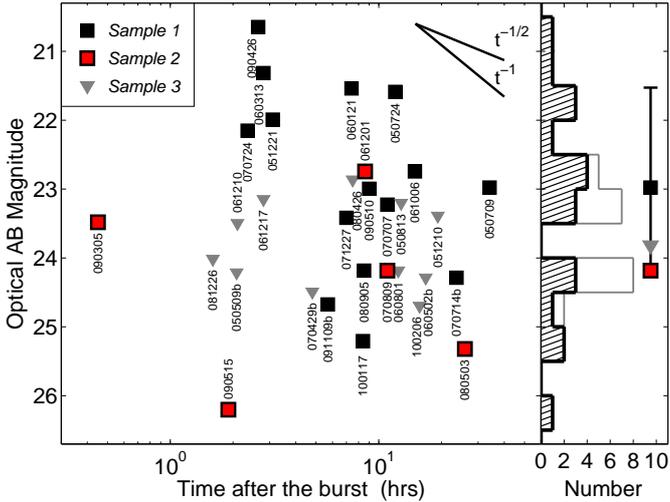}
\caption{Optical afterglow brightness on timescales of a few hours
after the burst for short GRBs with detected afterglows ({\it Sample
1}: black squares; {\it Sample 2}: red squares) or upper limits (gray
triangles).  The lines at the top right indicate the fading tracks for
afterglow decay rates of $\alpha=-0.5$ and $-1$.  The right panel
shows the projected histogram for the bursts with detected afterglows
(hatched) and upper limits (open).  The symbols mark the mean for each
sample, and the vertical bar marks the standard deviation for {\it
Sample 1}.  From \citet{ber10}.}
\label{fig:optag}
\end{figure}

In terms of the afterglow properties, the bursts in {\it Sample 2}
have a median optical brightness that is 1.4 mag fainter than the
bursts in {\it Sample 1} (Figure~\ref{fig:optag}, and X-ray fluxes at
8 hours that are about a factor of 2 times smaller.  Their
$\gamma$-ray fluences are similarly smaller, by about a factor of 5
\citep{ber10}.  The differences in optical afterglow brightness can be
due to lower densities or higher redshifts for {\it Sample 2}.  Both
of these scenarios would also explain the lack of bright coincident
hosts, since low densities may be indicative of large offsets and high
redshifts will lead to fainter host galaxies.  The difference in X-ray
brightness does not significantly constrain these two possibilities,
while the fainter $\gamma$-ray emission of the bursts in {\it Sample
2} points to high redshift as the likely explanation, since in the
context of the standard GRB model the prompt emission fluence does not
depend on the circumburst density.

I next turn to an analysis of the large-scale environments of the
bursts in {\it Sample 2}, particularly in comparison to the hosts of
bursts in {\it Sample 1}.  None of the 5 bursts have coincident hosts
to significantly deeper limits than the hosts in {\it Sample 1}.  I
therefore investigate the possibility of large offsets through the
calculation of chance coincidence probabilities for nearby galaxies,
as well as the possibility of a high redshift origin.  The chance
coincidence probability for nearby galaxies depends on both their
apparent magnitude and their distance from the optical afterglow
position.  The expected number density of galaxies brighter than a
measured magnitude, $m$, is \citep{hpm+97,bkd02,bsk+06}:
\begin{equation}
\sigma(\le m)=\frac{1}{0.33\times {\rm ln}(10)}\times
10^{0.33(m-24)-2.44} \,\,\,\,{\rm arcsec}^{-2},
\label{eqn:gal}
\end{equation}
and the chance coincidence probability for a given separation, 
$P(<\delta R)$, is then:
\begin{equation}
P(<\delta R)=1-{\rm e}^{-\pi (\delta R)^2\sigma(\le m)},
\label{eqn:prob}
\end{equation}
where for offsets substantially larger than the galaxy size, $\delta
R$ is the appropriate radius in Equation~\ref{eqn:prob} \citep{bkd02}.

\begin{figure}[ht!]
\centering
\includegraphics[angle=0,width=3.5in]{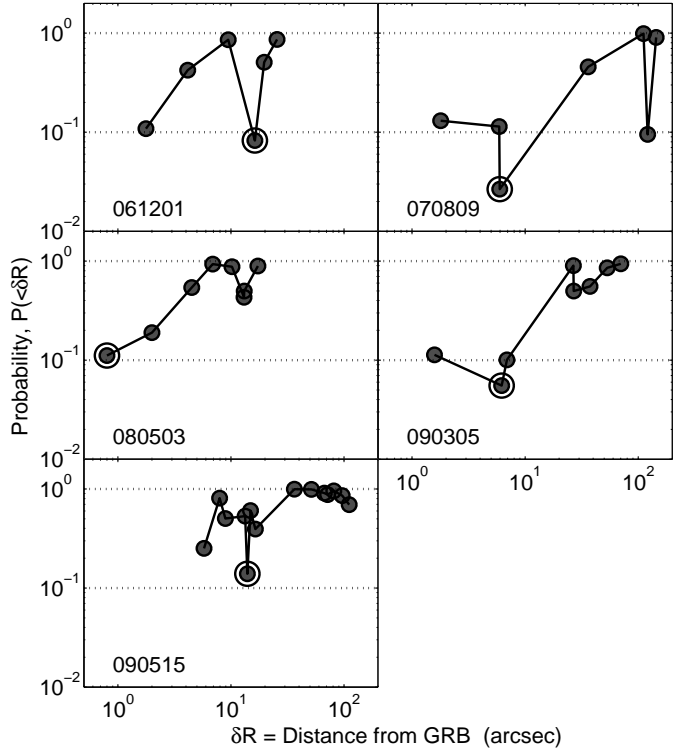}
\caption{Probability of chance coincidence as a function of distance
from a short GRB optical afterglow position for galaxies near the
location of each burst.  These are the galaxies marked in
Figure~\ref{fig:nohost}.  In each panel I mark the galaxy with the
lowest probability of chance detection with a circle.  In 4 of the 5
cases, the lowest probability is associated with galaxies that are
offset by $\sim 5-15\arcsec$.  Moreover, even the nearest galaxies are
offset by $\approx 1.6-5.8\arcsec$.  From \citet{ber10}.
\label{fig:prob1}} 
\end{figure}

The resulting distributions for each field are shown in
Figure~\ref{fig:prob1}; I include all galaxies that have probabilities
of $\lesssim 0.95$.  I find that for 4 of the 5 bursts, faint galaxies
($\sim 25-26$ mag) can be identified within $\approx 1.6-2\arcsec$ of
the afterglow positions, with associated chance coincidence
probabilities of $\approx 0.1-0.2$; in the case of GRB\,090515 I do
not detect any such faint galaxies within $\approx 5\arcsec$ of the
afterglow position.  For GRB\,080503 I also include the galaxy at an
offset of $0.8\arcsec$ and $m_{\rm AB}({\rm F606W})=27.3\pm 0.2$ mag
identified by \citet{pmg+09} based on their deeper stack of HST
observations.  On the other hand, for 4 of the 5 bursts I find that
the galaxies with the lowest probability of chance coincidence,
$\approx 0.03-0.15$, are brighter objects with offsets of about
$6-16\arcsec$ from the burst positions; only in the case of
GRB\,080503 the lowest chance coincidence is associated with the
nearest galaxy (see \citep{pmg+09}).

The use of {\it a posteriori} probabilities to assign {\it unique}
galaxy associations is fraught with difficulties.  First, for a given
apparent brightness, galaxies located further away from the GRB
position, potentially due to larger kicks and/or longer merger
timescales in the NS-NS merger framework, have higher probabilities of
chance coincidence.  Since no {\it a priori} model-independent
knowledge is available for the range of possible kicks and merger
timescales, it is not possible to rule out galaxies at very large
offsets for which $P(<\delta R)\sim 1$.  Indeed, a reasonable
constraint of $v_{\rm kick}\lesssim 10^3$ km s$^{-1}$ and $\tau_{\rm
merger}\lesssim 10$ Gyr leads to only a weak constraint on the offset
of $\lesssim 10$ Mpc.  At $z=0.1$ ($z=1$) this corresponds to about
$1.5^\circ$ ($0.3^\circ$), a projected distance at which nearly all
galaxies will have a chance coincidence probability of order unity.
However, this fact only means that we cannot rule out offsets that are
even {\it larger} than inferred from the most likely host association.

A second difficulty is that I am using angular offsets, which ignore
the potential wide range of redshifts (and by extension also
luminosities) of the various galaxies.  For example, if the faint
galaxies with small offsets are located at $z\gtrsim 1$, the
corresponding physical offsets are $\sim 15$ kpc, while if the
galaxies at $\sim 10\arcsec$ offsets are located at $z\sim 0.3$, the
offsets are only somewhat larger, $\sim 30$ kpc.  A galaxy at an even
lower redshift, $z\sim 0.1$, with an offset of 50 kpc will be located
about $30\arcsec$ from the GRB position and incur a large penalty in
terms of chance coincidence probability.  It is important to note,
however, that galaxies at lower redshift will generally have brighter
apparent magnitudes, partially compensating for the larger angular
separations (Equations~\ref{eqn:gal} and \ref{eqn:prob}).  In only a
single case (GRB\,070809) I find a galaxy with $P(<\delta R)\lesssim
0.1$ at $\delta R\gtrsim 1'$ (which at $z=0.043$ for this galaxy
corresponds to a physical offset of about 100 kpc).

A final complication, which is not unique to this subset of events, is
that only projected offsets can be measured, $\delta R=\delta R_{\rm
3D}\times {\rm cos}(\theta)$.  The measured offsets can be used as
lower limits on the actual offsets, while for the overall distribution
it is possible to apply an average correction factor of $\pi/2$, based 
on the expectation value for the projection factor, ${\rm cos}(\theta)$.

Despite these caveats, it is possible to address the probability that
{\it all} of the associations are spurious.  This joint probability is
simply the product of the individual probabilities \citep{bkd02}.  For
the faint galaxies at small angular separations the probability that
all are spurious is $P_{\rm all}\approx 8\times 10^{-5}$, while for
the galaxies with the lowest probability of chance coincidence the
joint probability is nearly 30 times lower, $P_{\rm all}\approx
3\times 10^{-6}$.  Conversely, the probabilities that {\it none} of
the associations are spurious are $\approx 0.42$ and $\approx 0.59$,
respectively.  These values indicate the some spurious coincidences
may be present in {\it Sample 2}.  Indeed, the probabilities that 1,
2, or 3 associations are spurious are $[0.40,\,0.15,\,0.027]$ and
$[0.34,\,0.068,\,0.006]$, respectively.  These results indicate that
for the faint galaxies it is not unlikely that $2-3$ associations (out
of 5) are spurious, while for the brighter galaxies $1-2$ associations
may be spurious.  This analysis clearly demonstrates why a joint
statistical study is superior to case-by-case attempts to associate
short GRBs with galaxies at substantial offsets.

Based on the possibility of association with the galaxies at
separations of $\sim 10''$, I obtained redshift for three of these
galaxies \citep{ber10}, leading to a star forming galaxy at $z=0.111$
(GRB\,061201), an early-type galaxy at $z=0.473$ (GRB\,070809), and an
early-type galaxy at $z=0.403$ (GRB\,090515).  The fainter host at
separations of a few arcsec likely reside at $z\gtrsim 1$.  The
redshifts provide an indication of the physical projected offsets
(\S\ref{sec:kick}).

\subsection{Undetected Faint Hosts at High Redshift?}
\label{sec:highz}

The redshifts of the GRBs in {\it Sample 2} can be constrained based
on their detections in the optical band (i.e., the lack of strong
suppression by the Ly$\alpha$ forest).  The afterglow of GRB\,061201
was detected in the ultraviolet by the {\it Swift}/UVOT and it is
therefore located at $z\lesssim 1.7$ \citep{rvp+06}.  The remaining
four bursts were detected in the optical $g$- or $r$-band, and can
therefore be placed at $z\lesssim 3$ or $\lesssim 4.3$; see
\citet{ber10} for details.

\begin{figure}[ht!]
\centering
\includegraphics[angle=0,width=3.5in]{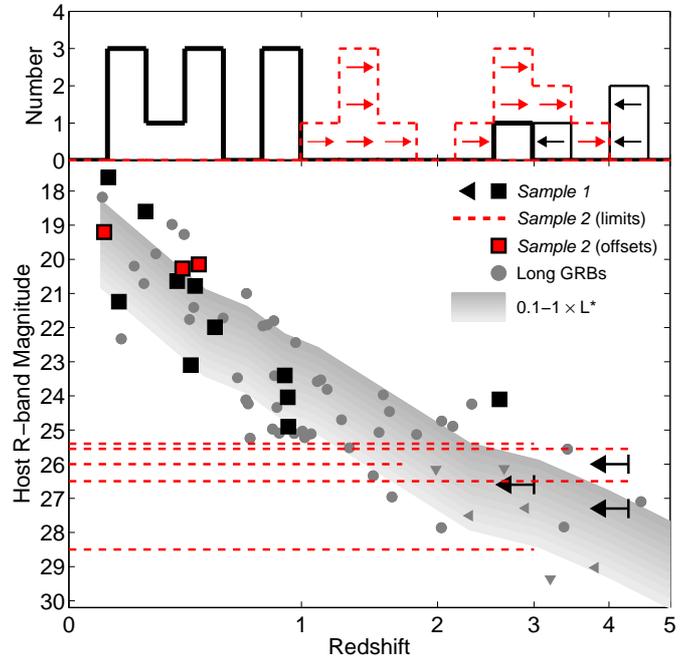}
\caption{{\it Main Panel:} Host galaxy optical magnitude as a function
of redshift for short GRB hosts (black squares), long GRB hosts
(detections: gray circles; non-detections or no redshifts: gray
triangles), and galaxies with a luminosity of $0.1-1$ L$^*$ (shaded
region).  The dashed lines mark the upper limits at the GRB positions
for the short GRBs with no coincident hosts.  The arrows mark the
upper limits on the redshifts of three bursts in {\it Sample 1} with
faint hosts, based on the detection of the afterglows in the optical
band (i.e., lack of a strong Lyman break).  If underlying host
galaxies exist for {\it Sample 2}, their non-detection indicates
$z\gtrsim 1.5$ (for $0.1$ L$^*$) or $\gtrsim 3$ (for L$^*$).  The
alternative possibility that they are located at similar redshifts to
the detected hosts, requires $\lesssim 0.01$ L$^*$, but this does not
naturally explain their fainter afterglows.  {\it Upper Panel:}
Projected redshift histogram for {\it Sample 1} (black) and {\it
Sample 2} (dashed red limits) under the assumption that the hosts are
$0.1$ L$^*$ galaxies ($z\sim 1.5$) and L$^*$ galaxies ($z\sim 3$).
From \citet{ber10}.
\label{fig:mags}} 
\end{figure}

It is possible to place additional constraints on the redshifts of any
underlying hosts using the existing sample short GRB host galaxies.
In Figure~\ref{fig:mags} I plot the $r$-band magnitudes as a function
of redshift for all available short GRB hosts from {\it Sample 1}.
For the faint hosts without known redshifts (GRBs 060121, 060313, and
070707), I place upper limits on the redshift using optical detections
of the afterglows.  A wide range of host magnitudes, $r_{\rm AB}\sim
16.5-27.5$ mag, is apparent.  I also plot the $r$-band magnitudes of
long GRB hosts \citep{sgl09}, as well as the $r-z$ phase space that is
traced by galaxies with luminosities of $L=0.1-1$ L$^*$.  I find
excellent correspondence between the hosts of long and short GRBs, and
the phase-space traced by $0.1-1$ L$^*$ galaxies, at least to $z\sim
4$.  In the context of these distributions, the available limits for
the short GRBs in {\it Sample 2} translate to redshifts of $z\gtrsim
1.5$ if they are 0.1 L$^*$ galaxies, or $z\gtrsim 3$ if they are L$^*$
galaxies.  The latter lower limits are comparable to the redshift
upper limits inferred from the afterglow detections.  I note that for
GRB\,080503, the limits of $\gtrsim 28.5$ mag and $z\lesssim 3$ from
the afterglow \citep{pmg+09} place even more stringent limits on the
luminosity of an underlying galaxy of $\lesssim 0.1$ L$^*$ galaxy.

The possibility that the five bursts originated at $z\gtrsim 3$ leads
to a bimodal redshift distribution (Figure~\ref{fig:mags}).  Nearly
all of the bursts in {\it Sample 1} with a known redshift (9/10) have
$z\approx 0.2-1$, with a median of $\langle z\rangle\approx 0.5$; the
sole exception is GRB\,090426 at $z=2.61$ \citep{adp+09,lbb+10}.  The
three bursts with faint coincident hosts have upper limits of
$z\lesssim 4$ from afterglow detections, while lower limit of
$z\gtrsim 1.5-2$ can be placed on these hosts if they have $L\gtrsim
0.1$ L$^*$.  Adding the {\it Sample 2} bursts with the assumption that
they have $z\gtrsim 3$ will furthermore result in a population of
short GRBs with a median of $z\sim 3$, and leave a substantial gap at
$z\sim 1-2$ (Figure~\ref{fig:mags}).  If the 5 bursts are instead
hosted by $0.1$ L$^*$ galaxies, the inferred lower limits on the
redshifts ($z\gtrsim 1.5$) lead to a potentially more uniform redshift
distribution.

It is difficult to explain a bimodal redshift distribution with a
single progenitor population such as NS-NS binaries, without appealing
to, for example, a bimodal distribution of merger timescales.  Another
possibility is two distinct progenitor populations, producing bursts
of similar observed properties but with distinct redshift ranges.
While these possibilities are difficult to exclude, they do not
provide a natural explanation for the short GRB population.

A final alternative explanation is that any underlying hosts reside at
similar redshifts to the known hosts in {\it Sample 1} ($z\sim 0.5$),
but have significantly lower luminosities of $\lesssim 0.01$ L$^*$.
This scenario would not naturally explain why the bursts in {\it
Sample 2} have fainter optical and X-ray afterglows, as well as lower
$\gamma$-ray fluences.  I therefore do not consider this possibility
to be the likely explanation.

\subsection{Large Offsets?}
\label{sec:kick}

While higher redshifts may explain the lack of detected hosts, the
fainter afterglows, and the smaller $\gamma$-ray fluences of the
bursts in {\it Sample 2}, this scenario suffers from several
difficulties outlined above.  The alternative explanation is that the
bursts occurred at significant offsets relative to their hosts, and
hence in lower density environments that would explain the faint
afterglow emission (though possibly not the lower $\gamma$-ray
fluences).  As demonstrated in the chance coincidence analysis, the
offsets may be $\sim 2\arcsec$ ($\sim 15$ kpc) if the bursts
originated in the faint galaxies at the smallest angular separations,
or $\sim 10\arcsec$ ($\sim 30-75$ kpc) if they originated in the
brighter galaxies with the lowest probability of chance coincidence
(Figure~\ref{fig:prob1}).

\begin{figure}[ht!]
\centering
\includegraphics[angle=0,width=3.5in]{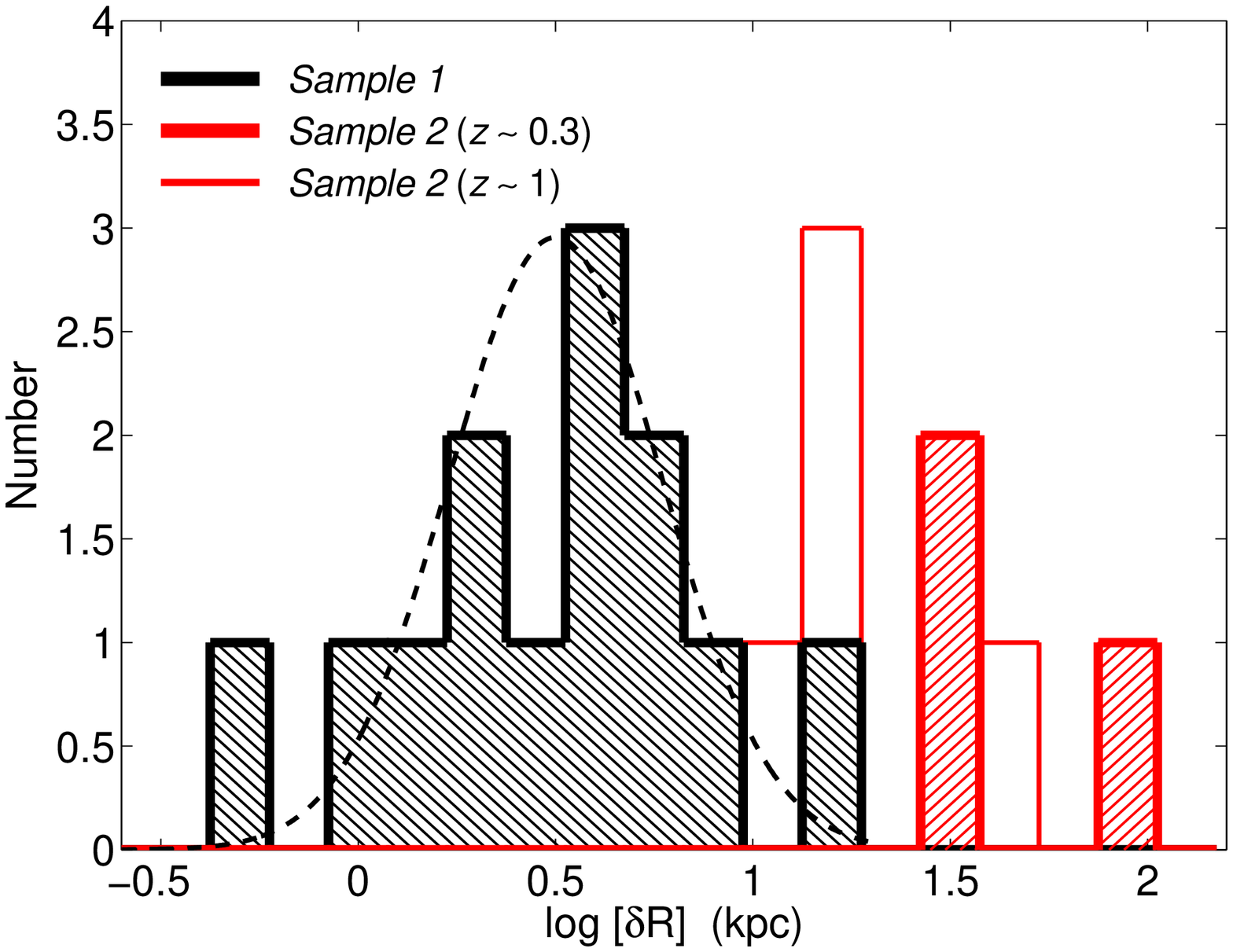}
\includegraphics[angle=0,width=3.5in]{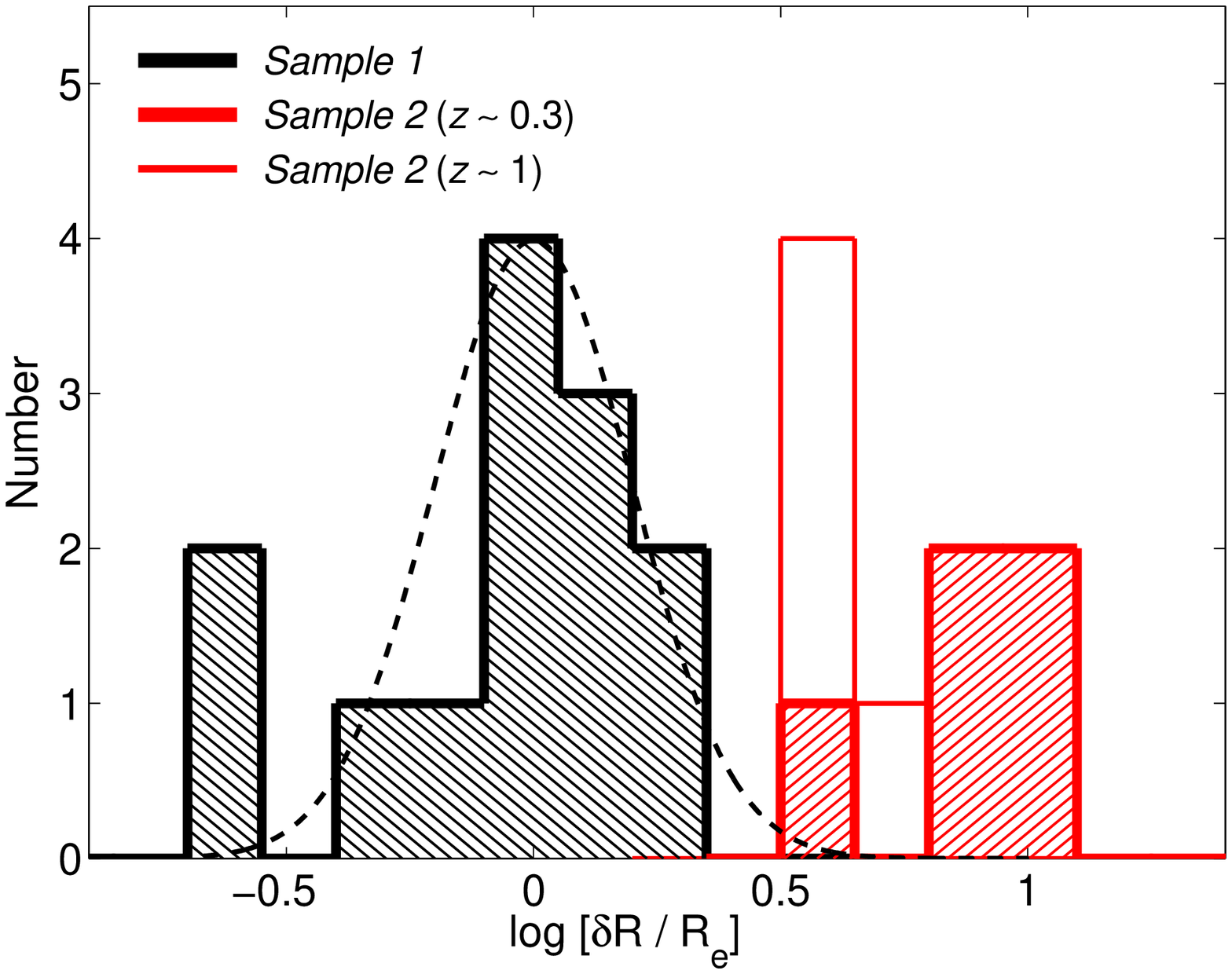}
\caption{{\it Top:} Histogram of projected physical offsets relative
to the host galaxy center for short GRBs with coincident hosts
(hatched black), and bursts with no coincident hosts if the galaxies
with lowest chance coincidence probability are the hosts (hatched
red), or if the faint galaxies with smallest angular separation are
hosts (open red); see Figure~\ref{fig:prob1}.  The dashed line is a
log-normal fit to the bursts with coincident hosts.  {\it Bottom:}
Same, but for offsets normalized relative to the host effective radii,
$R_e$.  The dashed line is a log-normal fit to the bursts with
coincident hosts.  From \citet{ber10}.
\label{fig:offset1}} 
\end{figure}

The projected physical offsets for {\it Sample 1} and {\it Sample 2}
are shown in Figure~\ref{fig:offset1}.  The mean and standard
deviation for {\it Sample 1} are $\delta R=4.2\pm 3.8$ kpc, and a
log-normal fit results in a mean of ${\rm log}(\delta R)\approx 0.5$
and a width of $\sigma_{{\rm log}(\delta R)}\approx 0.3$.  On the
other hand, the bursts in {\it Sample 2} have a mean offset of about
19 kpc if they arise in the faint galaxies with small angular
separation, or about 40 kpc if they arise in the brighter galaxies,
pointing to distinct distributions.

A similar result is obtained when considering the offset normalized by
each host's effective radius, $R_e$ (as advocated by \citet{fbf10}).
This quantity takes into account the varying sizes of the hosts due to
both intrinsic size variations and redshift effects.  It also gives a
better indication of whether the burst coincides with the host light
or is significantly offset.  As shown in Figure~\ref{fig:offset1}, the
host-normalized offsets of {\it Sample 1} have a mean and standard
deviation of about $1\pm 0.6$ $R_e$, and a range of about $0.2-2$
$R_e$.  A log-normal fit results in a mean of ${\rm log}(\delta
R/R_e)\approx 0$ and a width of $\sigma_{{\rm log}(\delta
R/R_e)}\approx 0.2$.  The bursts in {\it Sample 2} have much larger
host-normalized offsets, with $(\delta R/R_e)=7.3\pm 2.3$ if they
originated in the galaxies with the lowest chance coincidence
probability.  Even if I associate the bursts with the nearest faint
hosts, the distribution has a mean of about $4$ $R_e$, reflecting the
fact that the effective radii of the faint galaxies are smaller than
those of the brighter ones.

\begin{figure}[ht!]
\centering
\includegraphics[angle=0,width=3.5in]{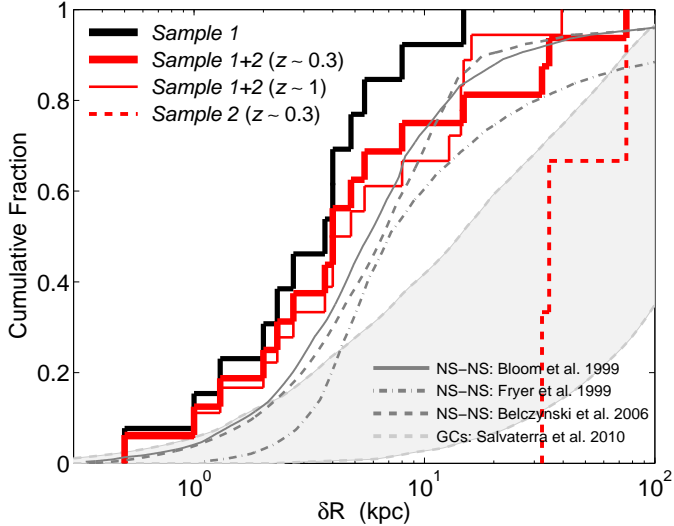}
\caption{Cumulative distributions of projected physical offsets for
short GRBs with coincident hosts (black line), and combined with
offsets for the hosts with the lowest probabilities of chance
coincidence (thick red line) or the faint hosts with smallest angular
offsets (thin red line).  Also shown are predicted distributions for
NS-NS kicks from several models \citep{bsp99,fwh99,bpb+06}, and for
dynamically-formed NS-NS binaries from globular clusters (shaded
region marks a range of predictions for host galaxy masses of $5\times
10^{10}-10^{12}$ M$_\odot$; \citep{sdc+10}).  The models with kick
velocities are in good agreement with the measured offset distribution
for either set of galaxy associations, while the globular clusters
model provides a poor match to the data.  From \citet{ber10}.
\label{fig:offset2}} 
\end{figure}

Thus, the distributions of physical, host-normalized, and angular
offsets exhibit a clear bimodality when associating the bursts in {\it
Sample 2} with the galaxies at $z\sim 0.1-0.5$.  The effect is still
apparent, though less pronounced in the case of association with the
faint galaxies at $z\gtrsim 1$.  Thus, if the offset scenario is
correct, the resulting distributions point to a possible bimodality
rather than a single continuous distribution of offsets.

The cumulative distributions of physical offsets for {\it Sample 1}
alone, and in conjunction with the two possible offset groups for {\it
Sample 2} are shown in Figure~\ref{fig:offset2}.  The combined
distributions have a median of about 4 kpc, driven by the bursts with
coincident hosts.  However, there is a clear extension to larger
physical offsets in the case of association with the brighter
galaxies, with about $20\%$ of all objects having $\delta R\gtrsim 30$
kpc.  The cumulative distributions are particularly useful for
comparison with NS-NS merger models since predictions exist for both
the kick scenario and the globular cluster origin model.

\section{Implications for the Progenitors}
\label{sec:conc}

The extensive analysis of host galaxy properties and the sub-galactic
environments of short GRBs presented above provides important insight
into the nature of the progenitors.  I address in particular the
popular NS-NS merger model, as well as delayed magnetar formation via
WD-WD mergers or WD accretion-induced collapse \citep{mqt08}.

\subsection{Host Galaxy Demographics and Properties}

The identified hosts of short GRBs with sub-arcsecond positions
(generally from optical afterglows) are dominated by star-forming
galaxies with a ratio of about 4:1, although I note that the nature of
the hosts of several short GRBs with sub-arcsecond positions remain
unknown mainly due to their faintness and/or lack of deep imaging and
spectroscopic observations.  Only if all the unidentified hosts are
early-type galaxies, would we have a ratio of 1:1.  The putative hosts
identified in coincidence with XRT error circles exhibit a similar
ratio of about 4:1.  This result is also supported by the
morphological analysis of short GRB hosts observed with {\it HST}
\citep{fbf10}, which indicates that the ratio of hosts with an
exponential disk profile versus a de Vaucouleurs profile is 4:1.
Thus, I conclude that the most conservative estimate of the ratio of
star forming to elliptical hosts is about 1:1, but that if the
well-studied (i.e., brighter) hosts are representative of the whole
sample, than the ratio is about 4:1.  A ratio of about 1:1 is expected
if short GRBs select galaxies by stellar mass alone.  Thus, the
existing demographics cannot rule out this scenario, but do appear to
point to an over-abundance of short GRBs in late-type galaxies,
possibly indicative of a short GRB rate that partially depends on star
formation activity.

Although star forming hosts appear to dominate the host population, it
is clear from a comparison of their luminosities, metallicities,
stellar masses, population ages, sizes, and star formation rates that
they are distinct from the hosts of long GRBs.  Namely, they exhibit
lower star formation activity and appear to be larger, more massive,
and dominated by more evolved stellar populations.  This indicates
that the short GRB progenitors are mostly related to the old stellar
populations within their hosts, and perhaps only partially to the
modest level of on-going star formation activity.  This is further
supported by the dearth of morphological galaxy merger signatures,
which point to delays of $\gtrsim 1$ Gyr relative to any
merger-triggered star formation episodes.  These conclusions are in
direct contrast to the massive star progenitors of long GRBs, which
select galaxies by star formation (and perhaps metallicity).  While
the relation to old stellar populations does not rule out models such
as WD-WD mergers or WD/NS AIC, it does disfavor young magnetars.  

As an aside, I note that this result demonstrates that caution should
be taken with the proposed re-classification of short and long GRBs
into Type I and II events, marking old and young progenitors,
respectively \citep{zzl+07,zzv+09}.  Such a new bimodal classification
may lead to the erroneous conclusion that short GRBs in late-type
galaxies (even if they track the on-going star formation activity)
share the same progenitors as long GRBs (e.g., \citep{vzo+09}) since
both would be classified as Type II.  At the very least, such a new
classification scheme may require a further breakdown of the Type II
events into those that result from massive stars versus those that
simply track star formation activity with a modest delay, e.g., Type
IIa and IIb.  Clearly, this is beyond the scope of the current short
GRB sample.

In the context of the NS-NS/NS-BH merger models, the host
demographics, coupled with the redshift distribution, provide a rough
constraint on the typical merger time delay
\citep{gp06,ngf06,bfp+07,zr07}.  In the formulation of \citet{zr07},
with a merger timescale probability distribution of $P(\tau)\propto
\tau^n$, the dominance of star forming hosts with a ratio of about 4:1
indicates $n\lesssim -1$.  I note that this is only a rough estimate
since the calculation is appropriate for $z\sim 0$, while the host
redshift distribution extends at least to $z\sim 1$.  Similarly, a
comparison of the observed redshift distribution (with a range of
$z\sim 0.2-2$ and a median of $\sim 0.7$) to the models of
\citet{ngf06} (their Figure 2), indicates a typical progenitor merger
timescale of about 3 Gyr (for a log-normal merger timescale
distribution).  A comparison to the models of \citet{gp06} (their
Figure 4), indicates a value of $n\sim -1$ (for a power law merger
timescale distribution), in good agreement with the results from host
demographics.

Thus, I conclude that based on the nature of short GRB host galaxies
and their redshift distribution, the merger timescale distribution for
the NS-NS/NS-BH coalescence models can be characterized by a power law
index of $\sim -1$ (power law model) or a typical value of $\sim 3$
Gyr (log-normal model).  {\it If} the subset of short GRBs in
late-type galaxies tend to track star formation activity, their
typical progenitor ages are instead $\sim 0.2$ Gyr \citep{lb10}.  It
should be noted that the merger timescale distribution for the
admittedly small sample of Galactic NS-NS binaries has been claimed to
be $\sim -1$ \citep{pir92,clm+04}.

\subsection{Short GRB Offsets}

The differential offsets measured from the {\it HST} observations
provide the most precise values to date for short GRBs, with a total
uncertainty of only $\sim 10-60$ mas ($\lesssim 1$ pixel),
corresponding to $\sim 30-500$ pc.  None of the offsets are smaller
than $\sim 1$ kpc, while this is the median offset for long GRBs.  On
the other hand, a substantial fraction of the offsets are only a few
kpc, indicating that at least some short GRBs explode within the
stellar component of their hosts (rather than their extended halos).
The median offset from the {\it HST} observations supplemented by
ground-based data is about 5 kpc (Figure~\ref{fig:offset}), roughly 5
times larger than for long GRBs.

As discussed above, the observed offset distribution is incomplete
since about $1/4-1/3$ of all short GRBs have only $\gamma$-ray
positions ($\sim 1-3'$), and a similar fraction have only XRT
positions, which generally lead to a range of $\sim 0-30$ kpc.  Taking
these limitations into account I find that the most robust constraints
on the offset distribution are that $\gtrsim 25\%$ of all short GRBs
have offsets of $\lesssim 10$ kpc, and that $\gtrsim 5\%$ have offsets
of $\gtrsim 20$ kpc.  In addition, for the current sample of of short
GRBs with sub-arcsecond afterglow positions and no coincident bright
hosts, I find evidence for offsets of $\sim 50$ kpc
(\S\ref{sec:kicks}).

Both the observed offset distribution and the various constraints are
in good agreement with predictions for the offset distribution of
NS-NS binaries in Milky Way type galaxies \citep{bsp99,fwh99,bpb+06}.
However, at the present a partial contribution from other progenitor
systems, such as delayed magnetar formation and even young magnetar
flares, cannot be ruled out.  The existence of modest ($\sim 10$ kpc)
and perhaps large ($\sim 50$ kpc) offsets in the sample suggests that
these latter models are not likely to account for {\it all} short
GRBs.

In the context of implications for the progenitor population, the
study of short GRB physical offsets by \citet{tko+08} led to the claim
that short GRBs with extended X-ray emission have systematically
smaller offsets than those with only a prompt spike, possibly due to a
systematic difference in the progenitors.  The {\it HST} sample of
\citet{fbf10} includes three short GRBs with significant extended
emission (050709, 050724, and 061006), and one burst (060121) with
possible extended emission ($4.5\sigma$ significance; \citep{dls+06}).
The physical offsets of these bursts are about 3.7, 2.7, 1.3, and 1
kpc, respectively, leading to a mean offset of about 2.2 kpc.  The
physical offsets of the bursts without extended emission, but with
precise afterglow positions (051221, 060313, and 061201) are 2.0, 2.3,
and 14.2 or 32.5 kpc, respectively.  The two events with no extended
emission and with XRT positions (050509B and 051210) have offsets of
about $54\pm 12$ and $28\pm 23$ kpc, respectively.  Including the
ground-based sample with optical afterglow positions, the bursts with
apparent extended emission (070714B, 071227, 080513, and 090510;
\citep{gcn6623,gcn7156,gcn9337,pmg+09}) have offsets of $\lesssim 4$,
14.8, $\sim 20$, and $\sim 5.5$ kpc, while the bursts without extended
emission (070724 and 070809) have offsets of 4.8 and $\sim 6.5$ kpc.
Thus, in the sample of events with sub-arcsecond positions, 6/8 bursts
with extended emission have offsets of $\lesssim 5$ kpc and 2/8 have
likely offsets of $\sim 15-20$ kpc.  In the sample without extended
emission, 4/5 have offsets of $\lesssim 6$ kpc and 1/5 has a likely
offset of $\sim 14-32$ kpc.  Thus, at the present it does not appear
that there is a significant difference in the two offset
distributions.

The inclusion of events with only XRT positions does not alter this
conclusion.  In particular, of the subset with no extended emission
only GRB\,050509B is likely to have a significant offset, while GRBs
051210, 060801, and 070429B have offsets ($28\pm 23$, $19\pm 16$, and
$40\pm 48$ kpc, respectively) that are consistent with zero.
Similarly, GRB\,061210 with extended emission has an offset of $11\pm
10$ kpc.  A continued investigation of the difference between short
GRBs with and without extended emission will greatly benefit from the
use of host-normalized offsets, which take into account the individual
hosts' effective radii.

\subsection{Large Offsets?}

The sample of short GRBs with optical afterglows represents about
$1/3$ of all short bursts, and may thus not be fully representative.
One often-discussed bias is that the bursts with optical afterglows
require a high circumburst density, and therefore have negligible
offsets.  However, from the analysis in \S\ref{sec:kicks}
\citep{ber10} it is clear that one explanation for the lack of
coincident hosts for the bursts in {\it Sample 2} is indeed large
offsets, {\it despite their detection in the optical band}.

As shown in Figure~\ref{fig:offset2}, the NS-NS merger model
predictions have a median of about 6 kpc, compared to about 4 kpc for
the observed sample.  On the other hand, the models predict $10-20\%$
of offsets to be $\gtrsim 30$ kpc, in good agreement with the observed
distribution in both the $\sim 15$ kpc and $\sim 40$ kpc scenarios
(\S\ref{sec:kicks}).  I note that the overall smaller offsets measured
from the data may be due to projections effects.  Indeed, the mean
correction factor of $\pi/2$ nicely reconciles the theoretical and
observed distributions.

In \S\ref{sec:kicks} I noted a bimodality in the physical and
host-normalized offsets for {\it Sample 1} and {\it Sample 2}
(Figure~\ref{fig:offset1}).  In the framework of NS-NS binary kicks
this bimodality may indicate that the binaries generally remain bound
to their host galaxies, thereby spending most of their time at the
maximal distance defined by $d_{\rm max}=2GM_{\rm host}/v_{\rm
kick}^2$ (i.e., with their kinetic energy stored as potential energy;
\citep{bpc+07}).  This would require typical kick velocities of less
than a few hundred km s$^{-1}$.

I further compare the observed offset distribution to predictions for
dynamically formed NS-NS binaries in globular clusters, with a range
of host galaxy virial masses of $5\times 10^{10}-10^{12}$ M$_\odot$
\citep{sdc+10}.  These models predict a range of only $\approx 5-40\%$
of all NS-NS mergers to occur within 10 kpc of the host center, in
contrast to the observed distribution with about $70\%$ with $\delta
R\lesssim 10$ kpc.  I stress that this result is independent of what
offsets are assigned to the bursts in {\it Sample 2} since they
account for only 1/4 of the bursts with optical afterglows.  On the
other hand, the globular cluster origin may account for the bimodality
in the physical and host-normalized offsets
(Figure~\ref{fig:offset1}), with the objects in {\it Sample 2} arising
in globular clusters and the objects with coincident hosts arising
from primordial NS-NS binaries.  This possibility also agrees with the
predicted fraction of dynamically-formed NS-NS binaries of $\sim
10-30\%$ \citep{gpm06}.  The cumulative offset distributions for {\it
Sample 2} alone (assuming the hosts are the galaxies with the lowest
probability of chance coincidence) is well-matched by the range of
predictions for dynamically-formed NS-NS binaries in globular clusters
(Figure~\ref{fig:offset2}).  In this scenario, however, the
implication is that short GRBs outside of globular clusters do not
experience kicks as expected for NS-NS binaries since the largest
measured offset is only 15 kpc.

Unless the populations of short GRBs with only X-ray or $\gamma$-ray
positions have fundamentally different offset distributions, I
conclude that the measured offsets of short GRBs and the predicted
offsets for NS-NS kicks are in good agreement, {\it if when treating
all short GRBs with optical afterglows as a single population}.
Alternatively, it is possible that the bimodal distributions of
physical and host-normalized offsets point to a progenitor bimodality,
with the bursts in {\it Sample 2} originating in globular clusters.

\subsection{Relation to the Host Galaxy Light Distribution} 

In addition to the projected offsets relative to their host centers,
it is apparent that short GRBs are more diffusely distributed relative
to their host light than long GRBs.  In particular, the locations of
short GRBs under-represent their overall host light distributions,
even in comparison to core-collapse and Type Ia SNe.  On the other
hand, it appears from the current small sample that short GRBs are
better tracers of their hosts' rest-frame optical light than UV light.
This result indicates that short GRBs arise in locations within their
hosts that trace the distribution of older stellar populations, and
clearly do not trace the sites of active star formation.  This result
provides strong support to the claim that although most short GRB
hosts are star forming galaxies, the bursts themselves are not related
to the star formation activity (\S\ref{sec:hosts}).

At the present, the sample of events with sufficiently precise
astrometry to determine the burst locations at the level of $\lesssim
1$ {\it HST} pixel is very small (6 events).  It is therefore not
possible to draw conclusions about the fraction of short GRBs that are
associated with old stellar populations as opposed to young
populations (as expected for young magnetars).  Luckily, there are at
least 10 additional events for which these measurements can be made
with future {\it HST} observations.  At the present, I conclude that
the stronger correlation of short GRBs with the rest-frame optical
light than UV light of their hosts is indicative of a dominant old
progenitor population.

\section{Conclusions}

While the sample of short GRBs with afterglow positions is still
significantly smaller than that of long GRBs, we have made significant
progress in understanding their galactic and sub-galactic
environments.  The results of host galaxy imaging and spectroscopy,
including high-resolution imaging with {\it HST}, point to an
association of short GRBs with old stellar populations within a range
of normal star forming and elliptical galaxies.  In nearly every
respect (star formation rates, metallicities, sizes, offsets, light
distribution) the environments of short GRBs are distinct from those
of long GRBs, indicating that their progenitors are not related to a
young progenitor population.

As I showed through the study of short GRB offsets, host galaxy
demographics, and the redshift distribution, the current observations
are fully consistent with NS-NS/NS-BH binary mergers.  However, a
partial contribution from other (mainly old) progenitor channels
(e.g., WD-WD mergers leading to magnetar formation, WD/NS AIC) cannot
be ruled out at the present.  Currently, we do not have conclusive
evidence for significant progenitor kicks, which are only expected in
the coalescence model.  Still, a few events with sub-arcsecond optical
positions do not directly coincide with bright host galaxies, and yet
reside within tens of kpc from bright, low-redshift galaxies.  This
may be suggestive of progenitor kicks, but it is also possible that
these bursts are associated with fainter hosts (likely at higher
redshift) with marginal offsets.

With continued vigilance, and a short GRB discovery rate of about 1
event per month, we are likely to gain further insight into the nature
of short GRB progenitors in the next few years, possibly with the
first detections (or significant limits) of gravitational waves.  As
argued in this review, host galaxy observations of existing and future
events will play a central role in our on-going quest to determine the
identity of short GRB progenitors.

\section{Acknowledgements}

I thank my collaborators on this work, D.~B.~Fox, E.~Nakar, W.~Fong,
S.~B.~Cenko, and A.~M.~Soderberg.  Observations used in this work were
obtained with the 6.5 meter Magellan Telescopes located at Las
Campanas Observatory, Chile, and the Gemini Observatory, which is
operated by the Association of Universities for Research in Astronomy,
Inc., under a cooperative agreement with the NSF on behalf of the
Gemini partnership: the National Science Foundation (United States),
the Science and Technology Facilities Council (United Kingdom), the
National Research Council (Canada), CONICYT (Chile), the Australian
Research Council (Australia), Minist?rio da Ci?ncia e Tecnologia
(Brazil) and Ministerio de Ciencia, Tecnolog?a e Innovaci?n Productiva
(Argentina).  Support for this work was provided by NASA/Swift Guest
Investigator grant NNX09AO98G, and by NASA/Chandra Award Number
GO9-0066X issued by the Chandra X-ray Observatory Center, which is
operated by the Smithsonian Astrophysical Observatory for and on
behalf of the National Aeronautics Space Administration under contract
NAS8-03060.


\end{document}